\documentclass[linenumbers]{pasj01}
\let\leftroot\relax
\let\uproot\relax

\usepackage{bm}
\usepackage{ulem}
\usepackage{amsmath}
\usepackage {afterpage}
\usepackage[switch,mathlines]{lineno}

\begin{document}

\title{New insights on the dynamics of  satellite galaxies: effects of the figure rotation of a host galaxy}
\author{Genta Sato, Masashi Chiba}
\altaffiltext{}{Astronomical Institute, Tohoku University, Aoba-ku, Sendai 980-8578, Japan}
\email{g.sato@astr.tohoku.ac.jp}


\maketitle

\begin{abstract}
We investigate a mechanism to form and keep a planar spatial distribution of satellite galaxies in the Milky Way (MW), 
which is called the satellite plane.
It has been pointed out that the $\Lambda$CDM cosmological model hardly explains the existence of such a satellite plane, so it is regarded as one of the serious problems in the current cosmology.
We here focus on a rotation of the gravitational potential of a host galaxy, i.e., so-called a figure rotation, following the previous suggestion that this effect can induce the tilt of a so-called tube orbit.
Our calculation shows that a figure rotation of a triaxial potential forms a stable orbital plane perpendicular to the rotational axis of the potential.
Thus, it is suggested that the MW's dark halo is rotating with its axis being around the normal line of the satellite plane.
Additionally, we find that a small velocity dispersion of satellites is required to keep the flatness of the planar structure, namely the standard derivation of their velocities perpendicular to the satellite plane needs smaller than their mean rotational velocity on the plane.
Although not all the MW's satellites satisfy this condition, some fraction of them called member satellites, which are prominently on the plane, satisfy it.
We suggest that this picture explaining the observed satellite plane can be achieved by the filamentary accretion of dark matter associated with the formation of the MW and a group infall of member satellites along this cosmic filament. 
  
\end{abstract}

\section{Introduction} \label{sec:intro}
It is known that the Milky Way (MW)'s satellite galaxies are aligned along a large-scale plane spatially and dynamically.
This planar structure is sometimes called Vast Polar Structure (VPOS) \citep{pawlowski2012}.
While it was thought that several classical satellites construct the planar structure \citep{pawlowski2012, pawlowski2020}, 
recent observations (e.g., with {\it Gaia}) show that more than half of the MW's dwarf satellites also distribute along the plane \citep{li2021, battaglia2022}.
Since the MW's satellite plane is nearly perpendicular to the stellar disk (see Table \ref{tab:obs}), it should be the independent structure of the MW's stellar system.

Since the $\Lambda$CDM model hardly explains statistically this anisotropic nature of the satellite plane (e.g., Kroupa et al. 2005), 
it is considered as one of the serious unsolved problems against the $\Lambda$CDM cosmology \citep{sales2022}.
This is called the satellite plane problem.
Some previous researches suggested that a massive satellite such as Large Magellanic Cloud (LMC) can raise the probability of forming a flattened distribution of other satellites in the MW to some extent (e.g., Samuel et al. 2021; Garavito-Camargo et al. 2021). 
However, the observation shows that other galaxies also have their satellite planes:
M31 \citep{koch2006, mcconnachie2006}, Centaurus A \citep{muller2016, kanehisa2023}, M81 \citep{chiboucas2013}, and M101 \citep{muller2017}.
Therefore, we expect that a satellite plane is not a structure occurring with a low probability due to a unique property of the MW, but a general structure based on a general mechanism of galaxy dynamics.

Here we focus on the rotational effect of the host halo which is called a figure rotation as a mechanism to form and maintain a satellite plane.
It has been known that a figure rotation of a triaxial potential affects the orbits of particles within it and stabilizes them on a tilted plane with regard to an axis of the potential \citep{habe1988, valluri2021}.
Some $\Lambda$CDM cosmological simulations show that a galaxy can gain a certain amount of pattern speed.
The distribution function of the pattern speed $\Omega_p$ has the median of $0.15h~{\rm km~s^{-1}~kpc^{-1}}$ and the width of $0.83h~{\rm km~s^{-1}~kpc^{-1}}$ ($h$ is the Hubble parameter in units of $100~{\rm km~s^{-1}~Mpc^{-1}}$) \citep{bailin2004} 
\footnote{The unit of ${\rm km~s^{-1}~kpc^{-1}}$ is nearly equal to ${\rm rad~Gyr^{-1}}$, as $1~{\rm km~s^{-1}~kpc^{-1}}=1.02~{\rm rad~Gyr^{-1}}$}.
Another previous research, \citet{bryan2007}, showed that $\Omega_p$ follows a log-normal distribution whose center is $0.2h~{\rm rad~Gyr^{-1}}$ and maximum value is $0.94h~{\rm rad~Gyr^{-1}}$ for a dark halo measured over $1h^{-1}~{\rm Gyr}$ by a $\Lambda$CDM simulation.
They also showed that the distribution evolves so that its center is $0.1h~{\rm rad~Gyr^{-1}}$ and maximum value is $0.24h~{\rm rad~Gyr^{-1}}$ over $5h^{-1}~{\rm Gyr}$.
\citet{valluri2021} showed that even such a slow pattern speed can change the trajectory of the Sagittarius stream to an observable extent.
Therefore, it is fully expected that a figure rotation causes a significant dynamic effect on its inner structure.

These previous researches also showed that a MW-like galaxy can possess a fast figure rotation.
When a galaxy has a figure rotation whose period is $0.2<P/({\rm Gyr})<5.0$
~(or angular speed is $1.23<\Omega_p/({\rm km~s^{-1}~kpc^{-1}})<30.0)$,
most box and tube orbits get unstable and chaotic.
On the other hand, a slower or faster figure rotation keeps stellar orbits stable \citep{deibel2011}. 
Another research \citep{carpintero2016} calculated the stability of a rotating galaxy whose period is $0.1<P/({\rm Gyr})<1.0$
(i.e., $6.14<\Omega_p/({\rm km~s^{-1}~kpc^{-1}})<61.4$).
They found that while such a fast figure rotation makes many (two-thirds) individual box and tube orbits chaotic, 
the whole rotating system is stable over the Hubble time.
These results provide important constraints on the MW's figure rotation.

However, these previous researches did not investigate the rotational effect on a scale of a satellite distribution over some Gyr and a few hundred kpc.
Thus, we explore the effect of a figure rotation on that scale and a structure formed by particles or satellites in a rotating gravitational potential.

We investigate the stable orbital structure (parent orbit) in a rotating potential.
In this calculation, we use fluidal particles because the fluidal terms have orbits of particles relax through the interactions, and shorten the time to reach their stable state. 
As a result, we can find a stable orbital plane easily and quickly.
Because this result, however, cannot be applied to the actual MW directly, we describe it in Appendix A.1.
Based on the result, we investigate the stability of the identified orbital plane with collisionless particles supposing actual satellites.
Since the orbit of a particle does not depend on its mass, the orbital plane of a fluidal particle is also the orbital plane of a collisionless particle.
However, the stability of the orbital plane should be strongly caused by the fluidal terms, so we investigate a condition to stabilize collisionless particles on the identified orbital plane.
Our calculation is performed over a long time, 10~Gyr, and over the radius that the MW's satellites reach, 200~kpc,
while few researchers have considered such a long-term orbital evolution of a galaxy over a large radius including its figure rotation.

This paper is organized as follows.
\S \ref{sec:obs_vpos} shows the current observed properties of the MW's satellite distribution.
\S \ref{sec:meth} explains our calculation method. 
\S \ref{sec:rslt2} shows our results using collisionless particles to investigate the stability of the orbital plane.
In \S \ref{sec:dscs}, we discuss the actual behavior of the observed satellites in a rotating halo.
\S \ref{sec:cncl} summarizes our work.
Additionally, Appendix A.1 shows the results using fluidal particles to explore a general stable orbital structure in a rotating potential, and A.2 explains details of our calculational condition; how to generate our initial particles.

\begin{figure*}[t]
\centering
\includegraphics[width=67mm]{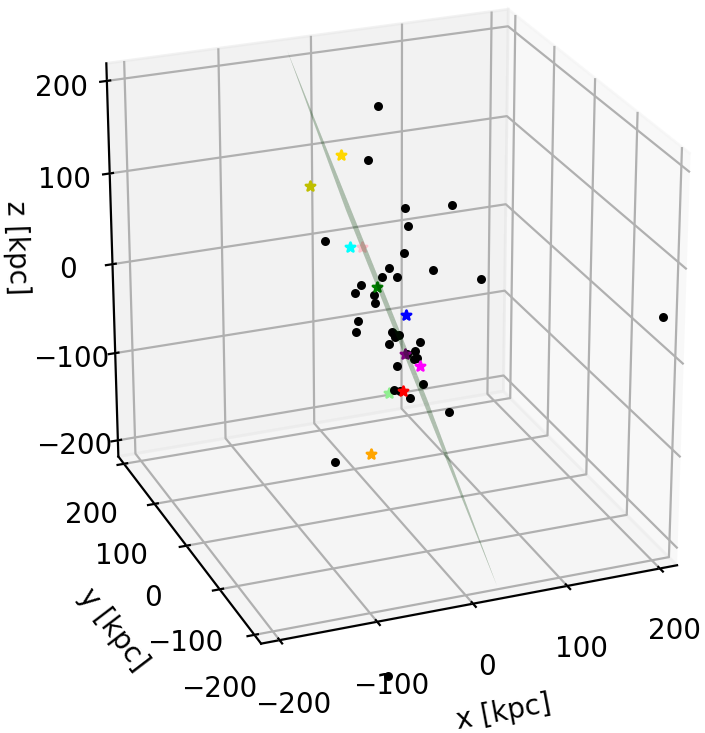}
\includegraphics[width=70mm]{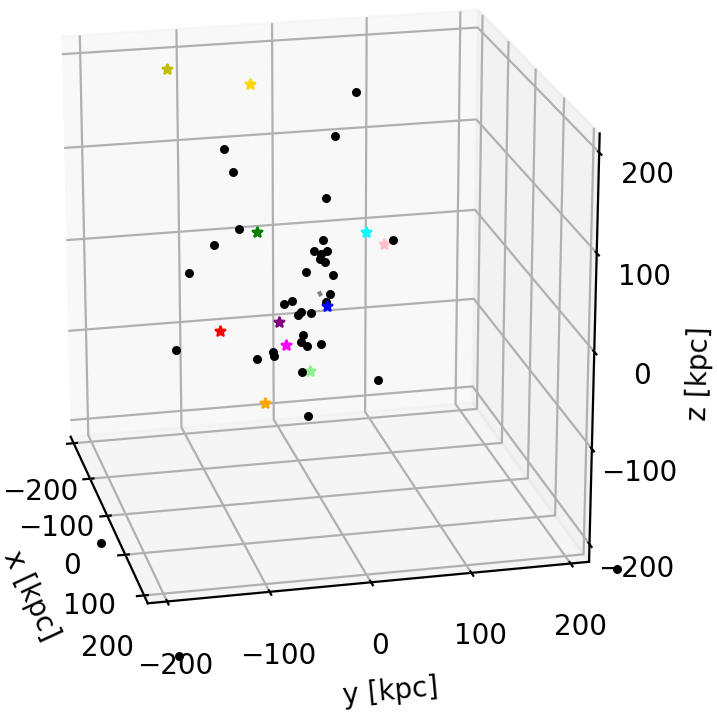}
\includegraphics[width=30mm]{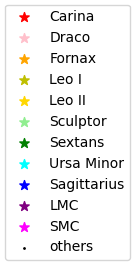}
\caption{The observed spatial distribution of the MW's satellites based on {\it Gaia} \citep{battaglia2022}.
Some famous classical satellites are colored (see the legend).
The left panel is the view from the edge-on direction of the fitted satellite plane (green line) determined in \S \ref{sec:obs_vpos}.
The right panel is from the face-on direction of the plane.
\label{fig:observed_satellite}}
\end{figure*}

\section{Observed Properties of the MW's Satellite Plane} \label{sec:obs_vpos}

\begin{figure*}[t]
\centering
\includegraphics[width=120mm]{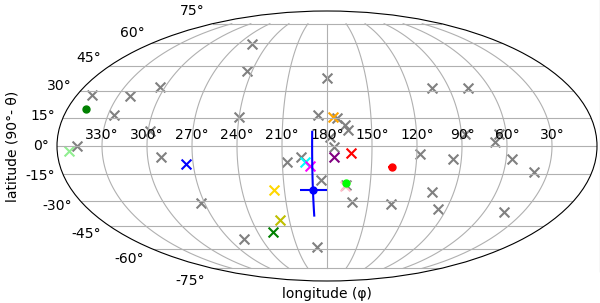}
\includegraphics[width=20mm]{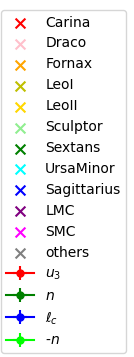}
\caption{Cross markers: the observed distribution of the orbital poles of the MW's satellites.
The classical satellites are colored as in Figure \ref{fig:observed_satellite}.
Round markers: the direction of three characteristic vectors of the MW's satellite distribution with uncertainties shown as error bars.
Because the sign of ${\bm n}$ is unimportant due to its definition,
its negative vector is also plotted.
\label{fig:observed_vec}}
\end{figure*}

The spatial distribution of the MW's satellites including faint dwarf satellites in the Galactocentric Cartesian coordinate 
\footnote{Assuming that the Sun is at (-8.129~kpc, 0, 0) \citep{gravitycollabo2018}, the circular velocity of the Sun is 229.0~${\rm km~s^{-1}}$ \citep{eilers2019} and the solar peculiar motion is (11.1, 12.24, 7.25)~${\rm km~s^{-1}}$ \citep{schonrich2010}} 
is shown in Figure \ref{fig:observed_satellite}.
The dynamical information of these satellites except for Sagittarius dwarf (Sgr), LMC, and Small Magellanic Cloud (SMC) are from the catalog of \citet{battaglia2022}.
The dynamics of Sgr, LMC, and SMC are referred from \citet{pawlowski2020}.
We extract satellite galaxies with available three-dimensional motions and adopt those being bound to the MW's gravitational field modeled as the oblate-like potential shown in \S \ref{subsec:eq}.
As a result, we adopt 49 satellites as our observational sample. 
Figure \ref{fig:observed_satellite} shows their distribution from the edge-on and face-on direction of the best-fitting plane defined as follows.

We adopt two methods to evaluate the properties of the spatial distribution of these satellites.
The first method uses an inertial tensor, which is defined as $S_{ij}=\sum_{k=0}^Nx_{i, k}x_{j, k}~(i,j=1,2,3)$, where $N$ is the number of satellites and $x_{i, k}$ is the $i$-th component of the position of the $k$-th satellite.
As usually used, the first, second, and third components are $x$, $y$, and $z$, in order.
$\{S_{ij}\}$ has the three eigenvalues $\lambda_1\geq\lambda_2\geq\lambda_3$ and the corresponding eigenvectors, ${\bm u}_1, {\bm u}_2$ and ${\bm u}_3$.
These vectors indicate the direction of the long, middle, and short axes, in order, of the ellipsoid which fits the satellite distribution best.
The three eigenvalues express the lengths of the corresponding axes of the fitted ellipsoid, 
so we use $\lambda_3/\lambda_1$ as the axial ratio of the satellite distribution.

In the second method, we fit the satellite distribution into a plane directly.
The equation of a plane in three-dimensional space is expressed as $a_0x+b_0y+c_0z=0$  ($a_0, b_0$, and $c_0$ are constants satisfying $a_0^2+b_0^2+c_0^2=1$).
The best-fitted plane is determined so that the summation of the square of the distances between the plane and each satellite is minimized with considering the observational uncertainties.
The distance between a plane and the $k$-th satellite is described as $d_k=|a_0x_k+b_0y_k+c_0z_k|$. 
Therefore, the definition of the best-fitted plane is what  $\sum_{k=0}^Nd_k^2/\sigma_k^2$ is minimized where $\sigma_k$ is the uncertainty of the position of the $k$-th satellite.
The mean of the physical distances from the plane defined as 
\begin{equation}
\Bar{d}\equiv\frac{1}{N}\sum_{k=1}^Nd_k, 
\label{mean_dist}
\end{equation}
indicates the thickness of the planar distribution.
The normal vector of this plane ${\bm n}$ is $(a_0, b_0, c_0)$.
When $\sigma_k=1$ for any $k$, ${\bm n}$ coincides with ${\bm u}_3$.

In the case of the distribution of the MW's satellites shown in Figure \ref{fig:observed_satellite},
we find that ${\bm u}_3$ points to $(\theta, \phi)=(101.60^{+0.70}_{-0.72}, 136.35^{+2.59}_{-2.50})~{\rm deg}$ where $(\theta, \phi)$ is the polar coordinate defined by the Galactocentric Cartesian coordinate.
On the other hand, ${\bm n}$ points to $(\theta, \phi)=(70.12^{+0.10}_{-0.10}, -12.92^{+0.10}_{-0.10})~{\rm deg}$.
It is shown that the MW's satellite plane is roughly perpendicular to the MW's disk plane no matter which vector is adopted as the direction of the satellite plane.

The axial ratio of the MW's satellite distribution is estimated as $\lambda_3/\lambda_1=0.367^{+0.016}_{-0.016}$.
On the other hand, the thickness of their distribution is $\Bar{d}=33.51^{+0.71}_{-0.69}~{\rm kpc}$.
As a comparison, the mean of the Galactocentric distance of satellites is $\Bar{r}_{\rm GC}=107.48^{+0.96}_{-0.89}~{\rm kpc}$, which is significantly larger than $\Bar{d}$.
Therefore, both the axial ratio and the thickness indicate that the MW's satellite forms a somewhat flattened distribution.

It is suggested that the satellite plane is not only the spatial structure but also the dynamical structure \citep{pawlowski2013, li2021}.
The cross marks in Figure \ref{fig:observed_vec} show the directions of the orbital poles of the MW's satellites.
Orbital poles are defined as the normalized angular momentum vector, ${\bm \ell}$.
As is shown, they roughly concentrate around $(\theta, \phi)=(90^\circ, 180^\circ)$. 
We estimate how and where they concentrate accurately.

The most concentrated point ${\bm \ell}_c$ is defined as follows.
At first, we describe the orbital pole of the $k$-th satellite as ${\bm \ell}_k$.
Next, we define the summation of the squares of angles between ${\bm \ell}_c$ and ${\bm \ell}_k$, i.e., 
\begin{equation}
\Delta^2=\frac{1}{N}\sum_{k=0}^N\arccos^2{({\bm \ell}_{c}\cdot{\bm \ell}_k)}. 
\label{mean_Delta}
\end{equation}
Finally, we seek ${\bm \ell}_c$ which minimizes $\Delta^2$ the most and we define this ${\bm \ell}_c$ as the most concentrated point. \citep{pawlowski2020}.
We find that ${\bm \ell}_c$ points to $(\theta, \phi)=(114.11^{+15.29}_{-32.33}, -170.08^{+8.82}_{-10.06})~{\rm deg}$, 
and $\Delta=83.52^{+1.94}_{-2.48}~{\rm deg}$.

\begin{table}[b]
\centering
 \caption{Summary of the parameters for the observed distribution of the MW's satellites}
  \begin{tabular}{|l|c|} \hline
    parameter & value or direction $(\theta, \phi)$\\ \hline \hline
    axial ratio $\lambda_3/\lambda_1$ & $0.367^{+0.016}_{-0.016}$ \\ \hline
    short axis ${\bm u}_3$ & $(101.60^{+0.70}_{-0.72}, 136.35^{+2.59}_{-2.50})~{\rm deg}$ \\ \hline
    thickness $\Bar{d}$ & $33.51^{+0.71}_{-0.69}$~kpc \\ \hline
    normal line ${\bm n}$ & $(70.12^{+0.10}_{-0.10}, -12.92^{+0.10}_{-0.10})~{\rm deg}$ \\ \hline
    concentration $\Delta$ & $83.52^{+1.94}_{-2.48}~{\rm deg}$ \\ \hline
    center of poles ${\bm \ell}_c$ & $(114.11^{+15.29}_{-32.33}, -170.08^{+8.82}_{-10.06})~{\rm deg}$ \\ \hline
  \end{tabular}
\label{tab:obs}
\end{table}

Finally, we show all the characteristic vectors by the round marks in Figure \ref{fig:observed_vec}.
It follows that three directions (${\bm u}_3$, $-{\bm n}$, and ${\bm \ell}_c$) are concentrated.
Considering that the sign of ${\bm n}$ is unimportant due to its definition, $-{\bm n}$ has the same physical meaning as ${\bm n}$.
Therefore, no matter how we define the satellite plane, it is almost perpendicular to the stellar disk.
The values of characteristic parameters and vectors are summarized in Table \ref{tab:obs}

\section{Method} \label{sec:meth}
In this work, we calculate the orbits of particles in a rotating potential to investigate the stability of flattened particle distribution with $N=1,000$ test particles.
The flattened distribution is supposed the observed satellite plane.
We use AGAMA software \citep{vasiliev2018} to treat the gravitational potential and calculation codes that we have written in C++ language to solve the equation of motion.
We calculate orbits of particles for $10~{\rm Gyr}$ and the timestep of the integration is $10^{-3}~{\rm Gyr}$.
When we carry out a simple calculation or draw a figure, we use Python language.

In the following, \S \ref{subsec:eq} explains the adopted gravitational potential and the equation of motion in detail and \S \ref{subsec:data} explains how to generate test particles as their initial state.

\subsection{Gravitational Potential and Equations of Motion}\label{subsec:eq}
We use a triaxial gravitational potential having a NFW-like density profile in its spherically symmetric limit as a host galaxy.
Its density profile is expressed as
\begin{align}
\rho({\bm x}) &= \frac{\rho_0}{\tilde{r}/r_s(1+\tilde{r}/r_s)^2}
\label{eq:triNFW}\\
{\rm where}~&{\tilde{r}}^2 =x^2 + (y/p)^2 + (z/q)^2
\label{eq:r_tilde}
\end{align}
and $p$ and $q$ are the axial ratios of isodensity ellipsoidal contours, $r_s$ is the scale length, and $\rho_0$ is the characteristic density with $\rho = \rho_0/4$ at $\tilde{r}=r_s$.
The gravitational potential $\Phi({\bm x})$ is calculated by the Poisson Equation using this mass density profile of a host galaxy.

For our calculation, we adopt two different sets of axial ratios, $(p, q)=(1.111, 0.889)$ and $(1.600, 0.837)$.
In both potentials, the long and short axes are along the y- and z-axis, respectively.
The former set of axial ratios is taken from the oblate-like potential given in \citet{valluri2021}. 
The latter set is aimed to reproduce the triaxial model for the MW potential in \citet{hayashi2007}, for which the corresponding density profile in its inner region has the axial ratios of short(z)/long(y)=0.72 and middle(x)/long(y)=0.78 in the isopotential contours. 
The axial ratios of the isopotential ellipsoids are summarized in Table \ref{tab:pote}.
We note that the former oblate-like potential is almost spherical.

\begin{table*}[bt]
\centering
 \caption{Summary of the axial ratios of the adopted host potentials. $p$ and $q$ are the parameters of the potentials as Equation \eqref{eq:r_tilde}.
 The columns with ``y/x'' show the ratios of the axial length along the y-direction over the x-direction of the isopotential ellipsoidal contours.
 The columns with ``z/x'' show the same but using z-direction instead of y-direction.
 The brackets indicate the Galactocentric radii measuring the axial ratios: ``$R_\odot$'' corresponds to 8.129~kpc and ``200'' is 200~kpc.}
  \begin{tabular}{|c|c|c|c|c|c|c|} \hline
    potential & $p$ & $q$ & y/x($R_\odot$) & z/x($R_\odot$) & y/x(200) & z/x(200)\\ \hline \hline
    oblate-like, almost spherical potential & 1.11 & 0.889 & 1.05 & 0.950 & 1.03 & 0.974\\ \hline
    MW-like, triaxial potential & 1.60 & 0.837 & 1.26 & 0.930 & 1.14 & 0.965\\ \hline
  \end{tabular}
\label{tab:pote}
\end{table*}

The normalizations for the density and scale radius are given as $\rho_0=8.5\times10^6~M_\odot~{\rm kpc}^{-3}$ and $r_s=19.6~{\rm kpc}$, respectively.
These values are determined so that the spherical potential assuming $(p, q) = (1, 1)$ coincides with the MW's halo presented by \citet{mcmillan2017}.

The motion of an $i$-th particle ($i=1-N$) is calculated by the following equations of motion, known as the equation in a rotating system;
\begin{align}
\frac{d{\bm x}_i}{dt} &= {\bm v}_i - {\bm \Omega} \times {\bm x}_i
\label{eq:EOM_x0}\\
\frac{d{\bm v}_i}{dt} &= -{\bm \nabla}\Phi({\bm x}_i) - {\bm \Omega} \times {\bm v_i}
\label{eq:EOM_v0}
\end{align}
where ${\bm x}_i$ and ${\bm v}_i$ are the position and velocity of the particle, respectively.

${\bm \Omega}=\Omega_c~(\sin{\theta_f}\cos{\phi_f}, \sin{\theta_f}\sin{\phi_f}, \cos{\theta_f})$ is the angular velocity of the figure rotation of the host potential.
We permit the direction of the rotation $(\theta_f, \phi_f)$ arbitrary. 
This direction should not necessarily be aligned with the coordinate axes, which are defined based on the configuration of the MW's stellar disk.
The concrete values of $(\theta_f, \phi_f)$ are determined for each initial particle distribution, explained in \S \ref{subsec:data}.

As the fiducial absolute value of $\Omega_c$ for the figure rotation, we set 0 or $1.0~{\rm km~s^{-1}~kpc^{-1}}$, where the latter corresponds to the period of $6.1~{\rm Gyr}$.
When we investigate the dependence of the result on the rotational speed, we also adopt $\Omega_c=0.5$ and $2.0~{\rm km~s^{-1}~kpc^{-1}}$, which correspond to the periods of $12$ and $3.1~{\rm Gyr}$, respectively.

\subsection{Initial State and Rotation Parameters} \label{subsec:data}
\begin{figure}[t]
\centering
\includegraphics[width=40mm]{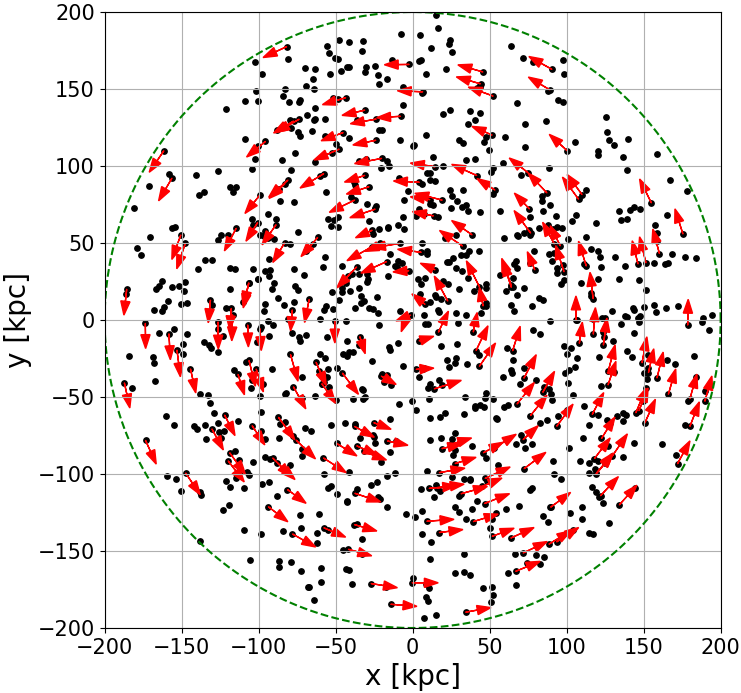}
\includegraphics[width=40mm]{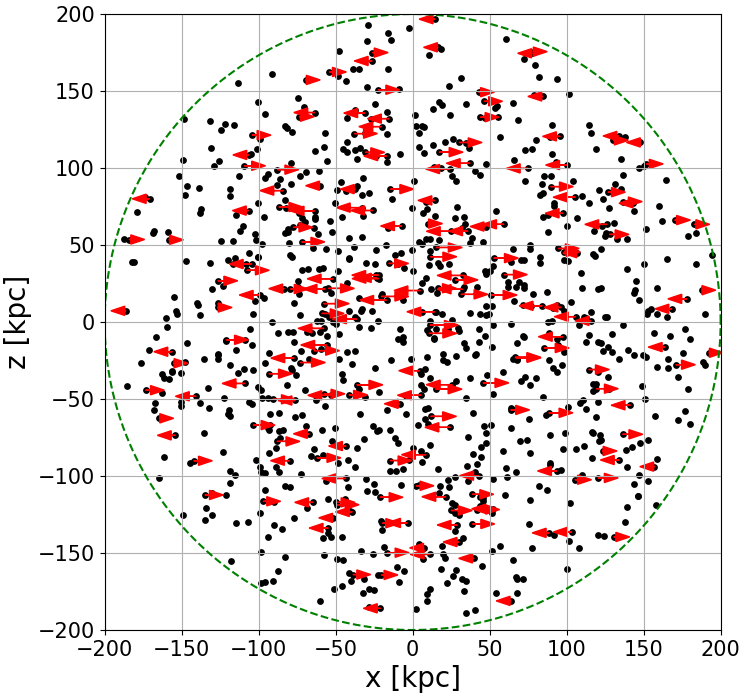}\\
\includegraphics[width=40mm]{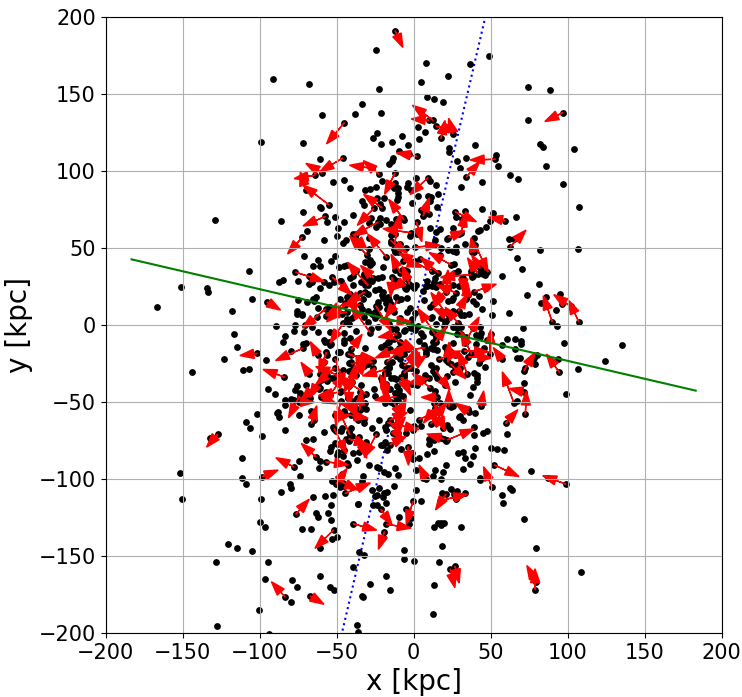}
\includegraphics[width=40mm]{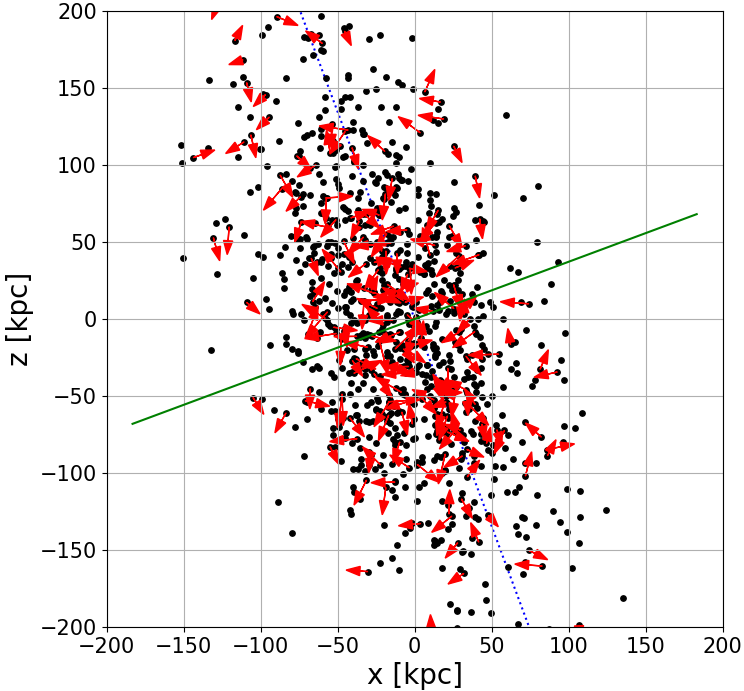}\\
\includegraphics[width=40mm]{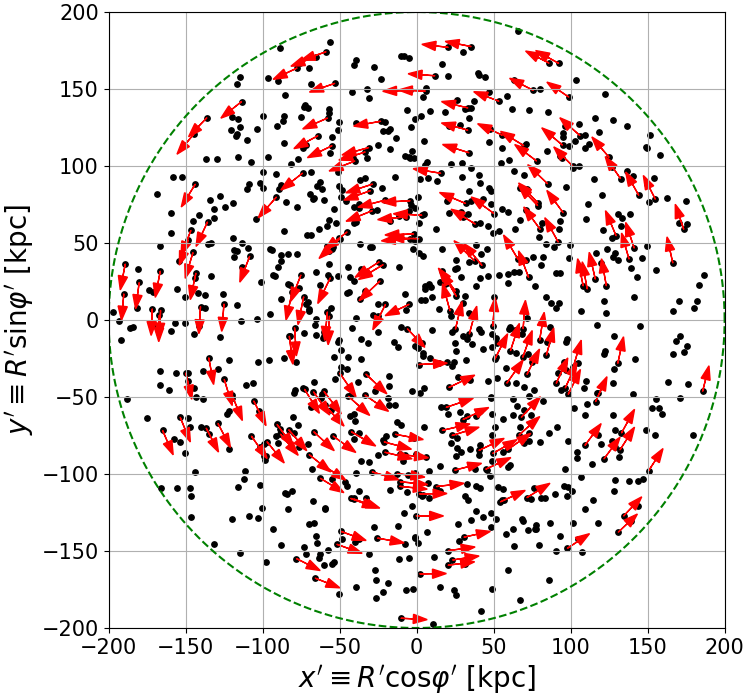}
\includegraphics[width=40mm]{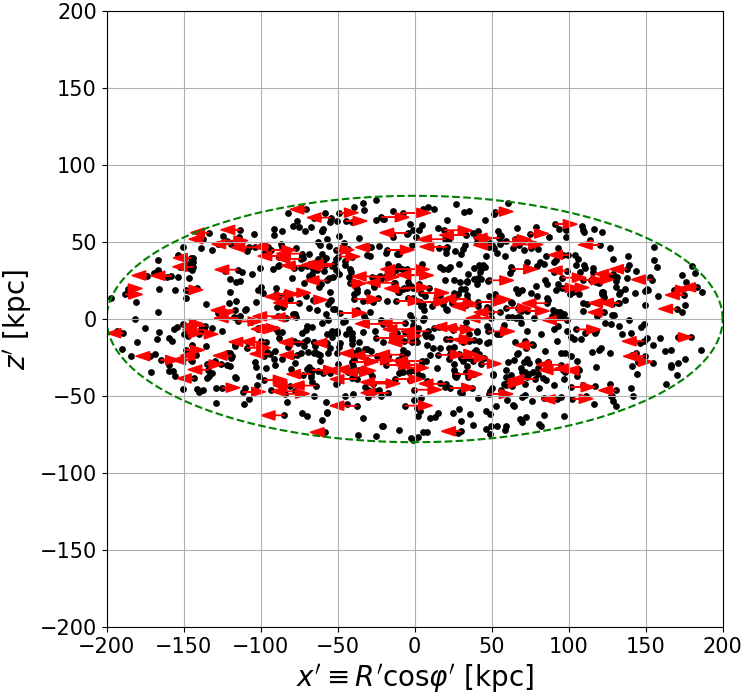}\\
\caption{Our initial particle distribution.
The panels from the top to bottom row show the first, second, and third initial states in order.
The left and right column shows the projection on the $x$-$y$ and $x$-$z$ plane in the first two rows or $x'$-$y'$ and $x'$-$z'$ plane ($x'\equiv R'\cos{\phi'},\ y'\equiv R'\sin{\phi'}$) in the lowest row.
We attach red arrows indicating the velocities to 100 particles chosen randomly.
The green dashed circles on the first row show the sphere with a radius of 200~kpc
and the green ellipses on the third row show the oblate ellipsoid with an axial ratio of 0.4 and a long axis of 200~kpc. 
The blue dotted line on the second panel shows the best-fitted plane on the $z=0$ or $y=0$ plane, 
and the green solid line is the normal line of the best-fitted plane, ${\bm n}$.
\label{fig:init}}
\end{figure}

We adopt three different initial states of the particle distribution.
The plots of these initial states are summarized in Figure \ref{fig:init}.
All initial states are particles with tube orbits around specific directions.

The first initial state is a sphere rotating around the z-axis.
This corresponds to particles with the short-axis tube orbit.
We place $N=1,000$ particles randomly and uniformly within a sphere whose radius is 200~kpc.
Their velocities are determined so that $v_R=v_z=0$ and the circular velocity is balanced with the centrifugal force from the host potential, i.e., $v_\phi=\sqrt{R(\partial\Phi/\partial R)}$ ($R\equiv\sqrt{x^2+y^2}$ is the radial coordinate on the equatorial plane).

In this case, we fix $\phi_f=90^\circ$ and assume $\theta_f=0^\circ,~30^\circ,~60^\circ,~90^\circ,~120^\circ, 150^\circ$, and $180^\circ$.
The axis of the figure rotation is on the plane made by the short-axis and long-axis of the host potential, i.e., the y-z plane.
The figure rotation with $\theta_f=0^\circ\ (180^\circ)$ corresponds to the picture that the potential has the same (reverse) rotation with the initial tube orbits.
The rotation with $\theta_f=90^\circ$ has the axis perpendicular to the rotational axis of the initial tube.

The second initial state is based on the observed flattened distribution of the MW's satellites shown in Figure \ref{fig:observed_satellite}.
This initial distribution of particles has a normal line with ${\bm n} = (\theta, \phi) = (70^\circ, -13^\circ)$, i.e., all the particles have the same rotational direction. 
This case can be associated with the hypothesis that all the satellites are accreted from the same direction.
The specific method to generate this second initial state is described in Appendix A.2 in detail.

In the second case, we assume $(\theta_f, \phi_f) = (0^\circ,~90^\circ),\\(30^\circ,~90^\circ),~(60^\circ,~90^\circ),~(90^\circ,~90^\circ)$, and $(70^\circ,~-13^\circ)$.
The fifth direction coincides with the normal vector of the best-fitted plane, ${\bm n}$.
Although we consider the pairs of the reverse figure rotation (e.g. $(\theta_f, \phi_f)=(30^\circ, 90^\circ)$ and $(150^\circ, 90^\circ)$) in the case of the first initial state,
we adopt only either of the pairs in this second initial state.
This is because we find that these pairs of the figure rotation cause almost the same results in the first initial state,  
so we understand that either of the pairs can be abbreviated from the symmetry.

The third initial state consists of tube orbits around $(\theta, \phi)=(70^\circ, -13^\circ)$.
We define a tilted oblate ellipsoid with an axial ratio of 0.4 and with its short axis pointing to $(\theta, \phi)=(70^\circ, -13^\circ)$, within which we set up $N=1,000$ particles randomly.
Here, we define a cylindrical coordinate $(R', \phi', z')$ based on ${\bm n}$.
The velocities of particles are determined so that the azimuthal component ($v_{\phi'}$) is balanced with the centrifugal force from the potential.
Other velocity components are given following a Gaussian distribution function whose mean is 0 and dispersion is the square of $k$ times the mean of the azimuthal velocity component, i.e., $\Bar{v}_{\alpha'}=0$ and $\sigma_{v_{\alpha'}}=k\Bar{v}_{\phi'} (\alpha=R$ or $z)$.
We adopt $k=0, 0.1, 0.3, 0.5, 0.7$, and 1.0.

In the third case, we fix $\theta_f=70^\circ$ and assume $\phi_f=-13^\circ, -2^\circ, 8^\circ, 19^\circ, 35^\circ, 51^\circ,$ and $ 85^\circ$.
These angles are determined so that the offset between the axes of the figure rotation and the short axis of the initial particle distribution is $0^\circ, 10^\circ, 20^\circ, 30^\circ, 45^\circ, 60^\circ$, and $ 90^\circ$, in order.


\section{Result: Stability of a Satellite Plane} \label{sec:rslt2}
In the following, we evaluate the particle distribution by its axial ratio $\lambda_3/\lambda_1$ and the direction of the short axis of the best-fitted ellipsoid ${\bm u}_3$ defined in \S \ref{sec:obs_vpos}.
As mentioned in \S \ref{sec:obs_vpos}, ${\bm u}_3$ coincides with the normal vector of the best-fitted plane ${\bm n}$ 
because the uncertainties of particle positions are not taken into account in this computational work.
However, since it is generally difficult to show and evaluate the evolution of the vector ${\bm u}_3$ quantitatively, we also use the angle, $A_p\equiv\arccos{({\bm \Omega}/\Omega_c\cdot{\bm u}_3)}$, i.e., the offset of ${\bm u}_3$ from the axis of the figure rotation.

In the fluidal simulations described in Appendix A.1, we have found a stable orbital plane in a rotating potential.
In an almost spherical potential, a stable plane is precessing around the axis of the figure rotation.
On the other hand, in a triaxial potential, a stable plane is perpendicular to the axis of the figure rotation, but this structure appears only if the offset between the directions of the figure rotation and the particle orbits is small.
Based on these results, we explore how the stable orbital plane formed by collisionless particles behaves, compared with the fluidal cases, whereby we will get further insights into the origin of the observed planar structure of satellite galaxies.

\begin{figure*}[t]
\centering
\includegraphics[width=80mm]{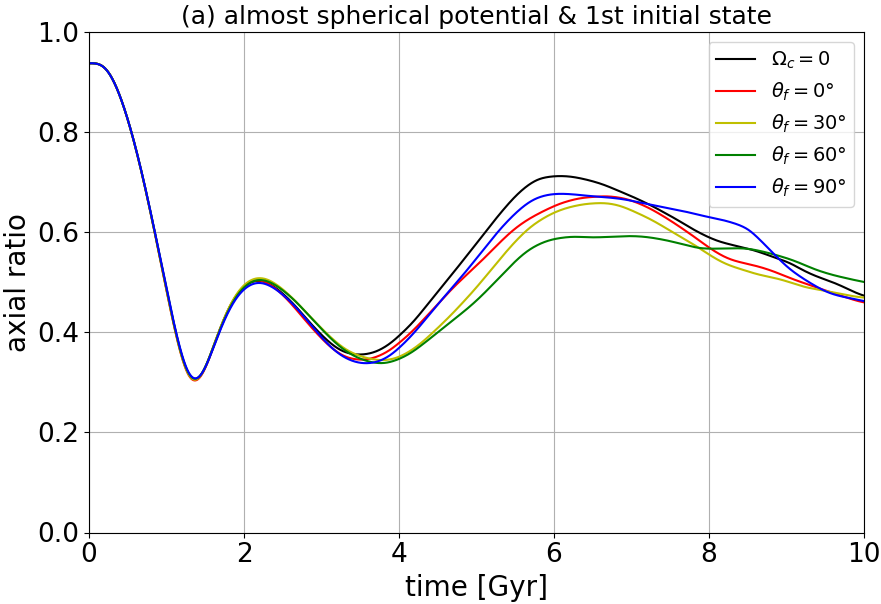}
\includegraphics[width=80mm]{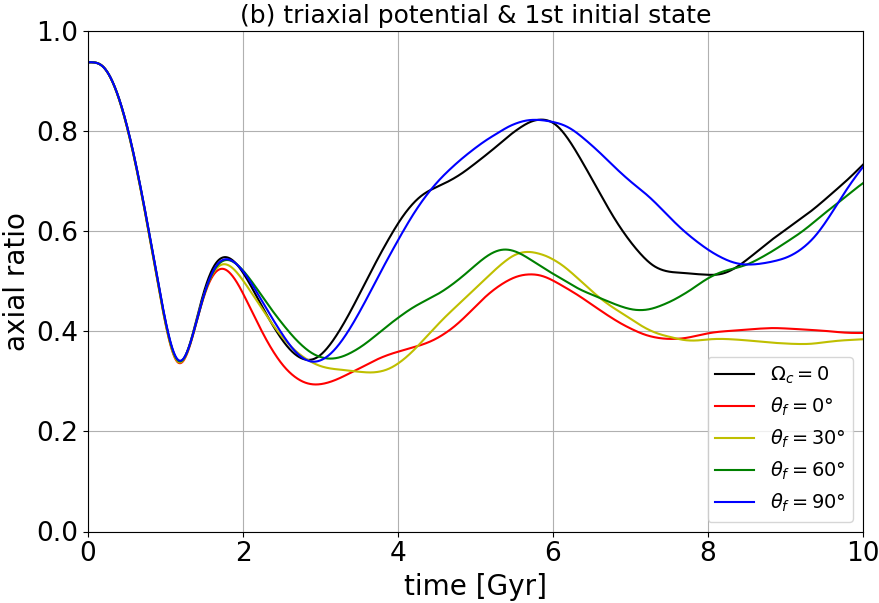}\\
\includegraphics[width=80mm]{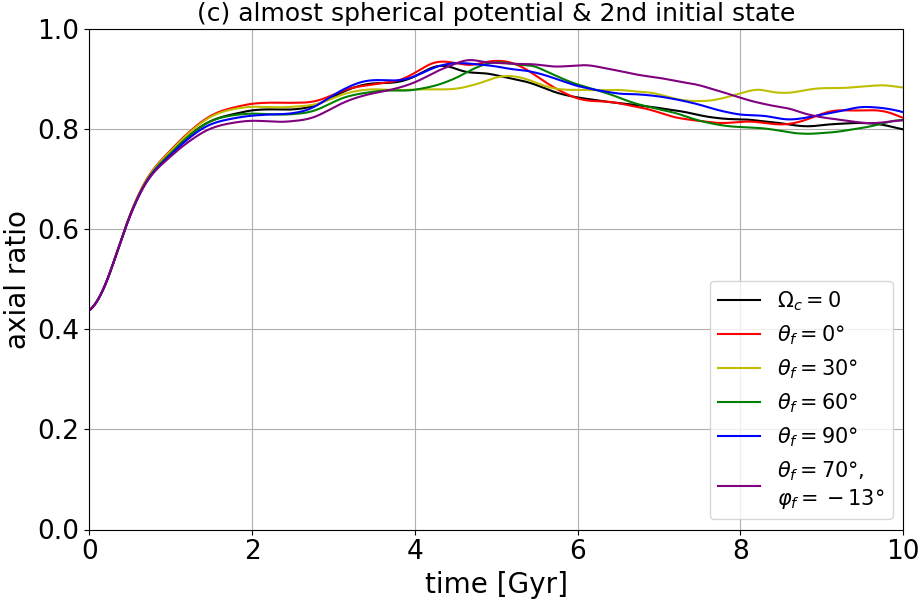}
\includegraphics[width=80mm]{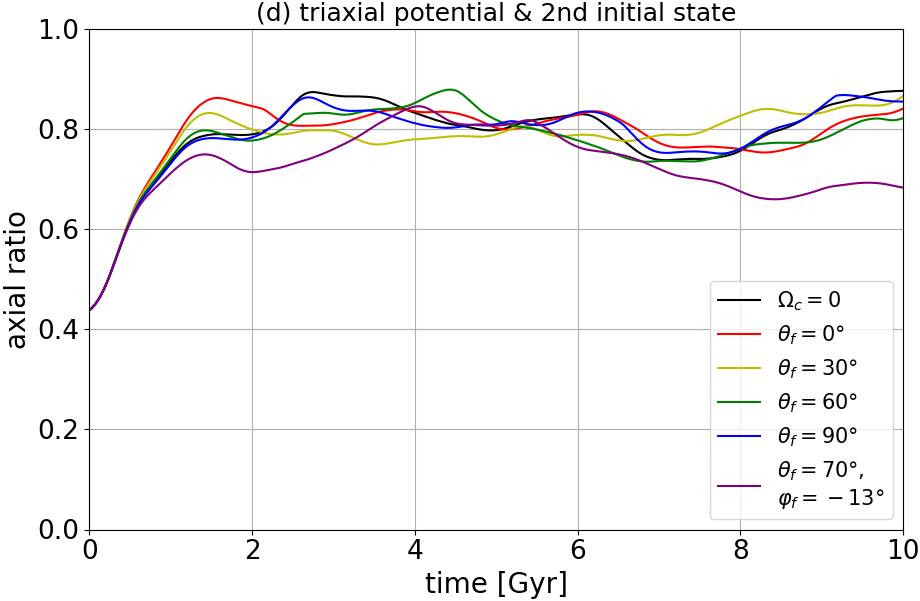}
\caption{The evolution of the axial ratio of the collisionless particle distribution.
The upper and lower rows show the result with the first and the second initial conditions, respectively.
The left and right columns show the cases with the almost spherical and the triaxial potential, respectively.
The black graph labeled ``$\Omega_c=0$'' is the case without the figure rotation.
The colored graphs show the cases with figure rotation with the direction labeled in the legends.
When a description about $\phi_f$ is omitted in a label, it expresses $\phi_f=90^\circ$ (the same applies to the following figures).
\label{fig:axisratio2}}
\end{figure*}

We carry out the orbital calculation for collisionless particles (supposing the actual satellites) with Equation \eqref{eq:EOM_x0} and \eqref{eq:EOM_v0}.
We omit the cases with $\theta_f>90^\circ$ because the fluidal calculation shows that their results can be inferred by the symmetry from $\theta_f<90^\circ$ cases.
Figure \ref{fig:axisratio2} shows the evolution of the axial ratio of collisionless particles.
It shows that the axial ratio is larger and less stable than the fluidal cases.
However, if the potential is triaxial and the offset between the directions of the figure rotation and the initial tube orbits is sufficiently small (the red and yellow lines in panel (b)), the axial ratio is settled into a small value of $\sim0.4$.
Although this axial ratio is as small as the observation, the setup for the potential and the initial condition is ideal, so we cannot simply apply this case to the actual MW's satellites.
Additionally, the exceptional case derived from the fluidal calculation ($(\theta_f, \phi_f)=(70^\circ, -13^\circ)$, the purple line in panel (d)), also has a relatively small axial ratio $\sim0.7$, but this is much larger than the case of fluidal particles (Figure \ref{fig:axisratio1}) and the observed satellite plane (Table \ref{tab:obs}).
These relatively small axial ratios do not appear in the almost spherical potential (panel (a) and (c)), thus, the flattened structure of simulated particles is a characteristic property in a rotating triaxial potential.

\begin{figure*}[t]
\centering
\includegraphics[width=80mm]{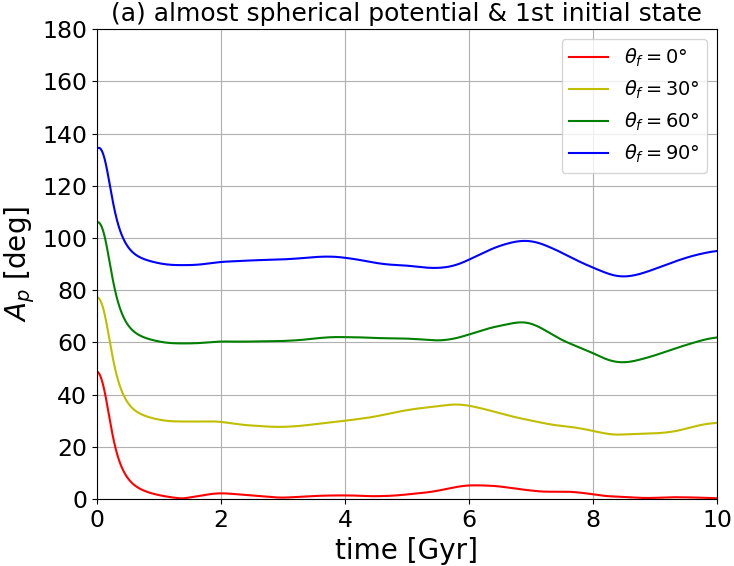}
\includegraphics[width=80mm]{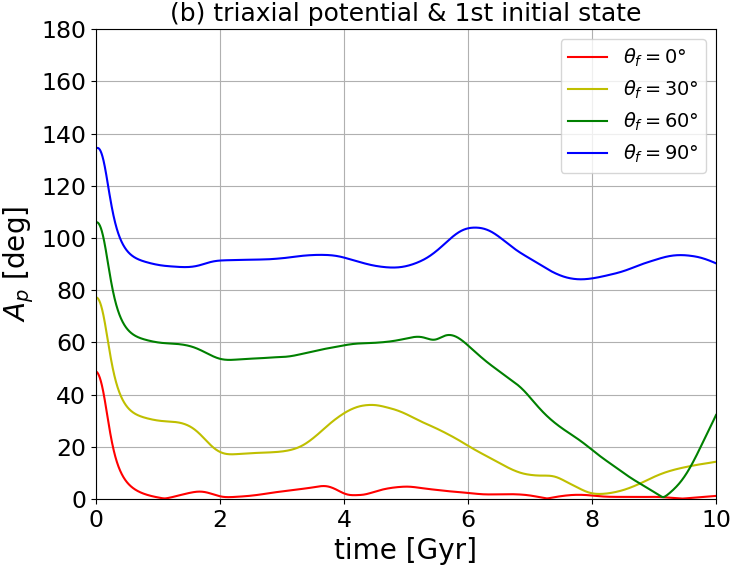}\\
\includegraphics[width=80mm]{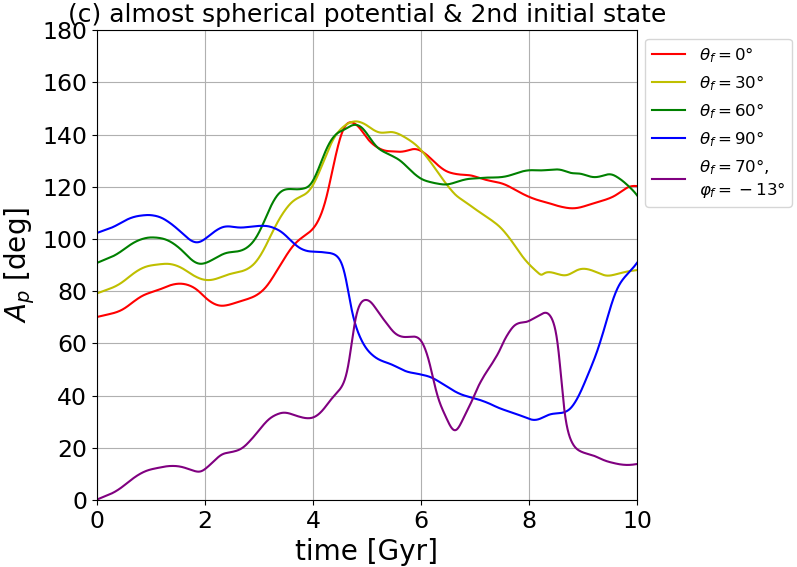}
\includegraphics[width=80mm]{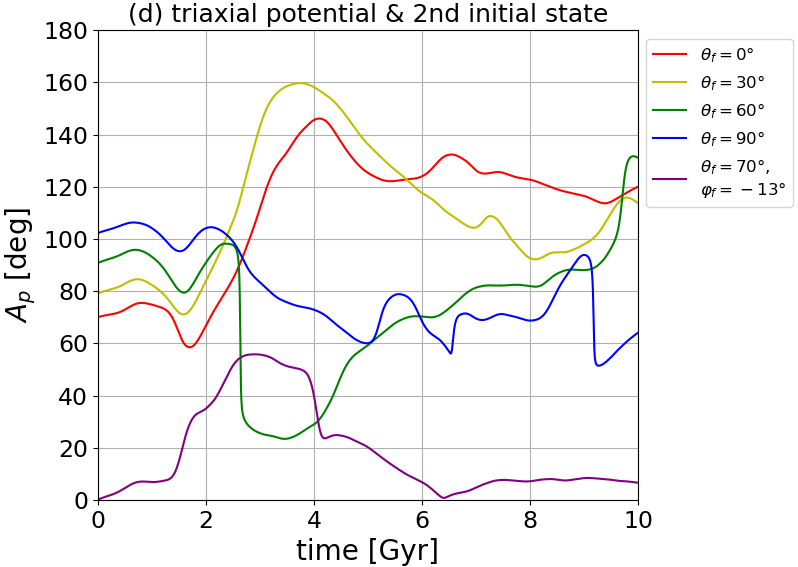}
\caption{The evolution of the direction of the particle distribution.
As in Figure \ref{fig:axisratio2}, the left and right columns show the cases with the almost spherical (i.e., oblate-like) and the triaxial (i.e., MW-like) potential, respectively.
The upper and lower rows show the result with the first and the second initial conditions, respectively.
\label{fig:amplitude2}}
\end{figure*}

Figure \ref{fig:amplitude2} shows the evolution of the particle distribution expressed by $A_p$.
As in the case of the fluidal particle case (Figure \ref{fig:amplitude1} in Appendix A.1), $A_p$ in panel (a) is nearly constant with time, coinciding with $\theta_f$, but slightly unstable in the long term $\gtrsim6~{\rm Gyr}$.
It means that the ellipsoidal distribution of particles is precessing around the axis of the figure rotation.
We also find that the frequency of the precession is equal to $\Omega_c$, as in the case using fluidal particles (Figure \ref{fig:freq10}).
Additionally, in panel (b), the tendency for $A_p$ to damp is more significant.
Namely, the ellipsoidal distribution of particles gets stationary so that its short axis coincides with the rotational axis of the potential.
On the other hand, panel (c) shows that $A_p$ changes drastically, which suggests that the direction of the particle distribution is so unstable.
In panel (d), $A_p$ in most cases is also unstable but the exceptional case mentioned above (purple line) shows a stable $A_p\sim0$ after 6~Gyr.
In summary, it is suggested that the relatively flattened structure of collisionless particles (red or yellow lines in panel (b) and purple line in panel (d) in Figure \ref{fig:axisratio2}) is nearly perpendicular to the rotational axis of the figure rotation.
To obtain further conditions on the state of the satellite plane so that it is as flattened as the observed one, 
we now consider the third initial state of collisionless particles given in \S \ref{sec:meth} for the case of a rotating triaxial potential.

\begin{figure*}[tp]
\centering
\includegraphics[width=55mm]{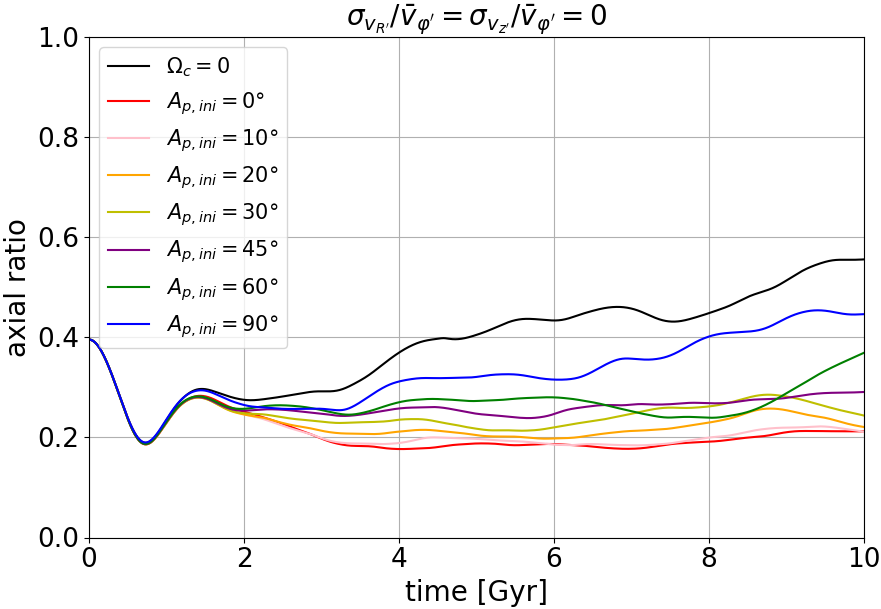}\\
\includegraphics[width=55mm]{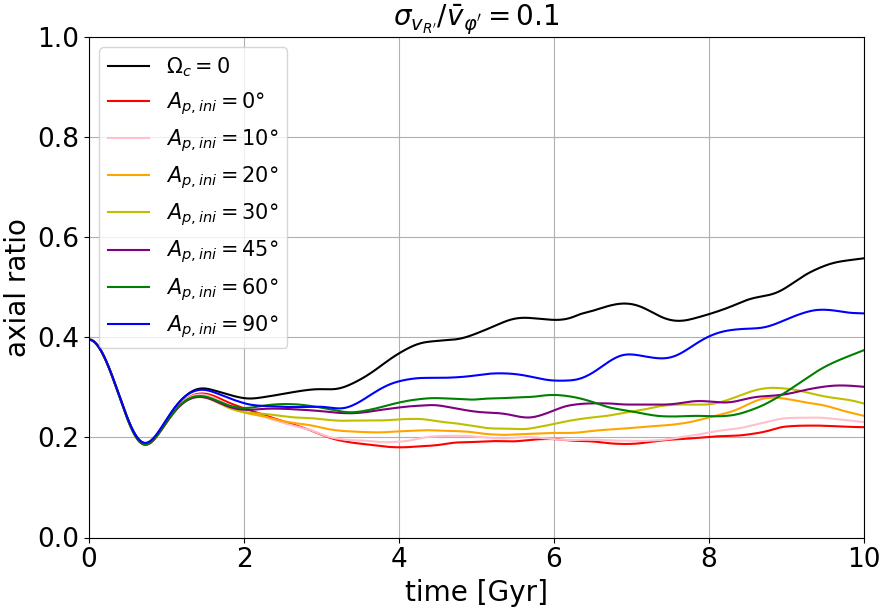}
\includegraphics[width=55mm]{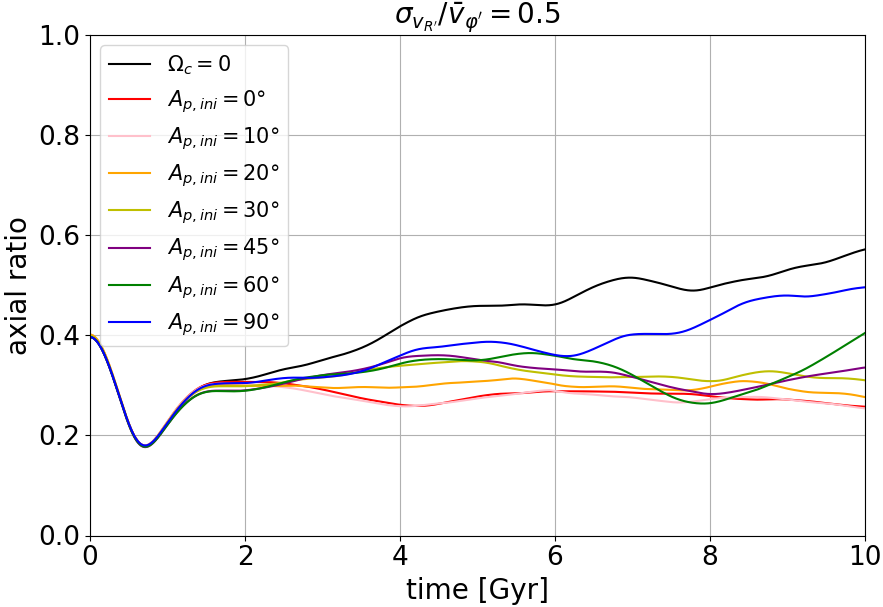}
\includegraphics[width=55mm]{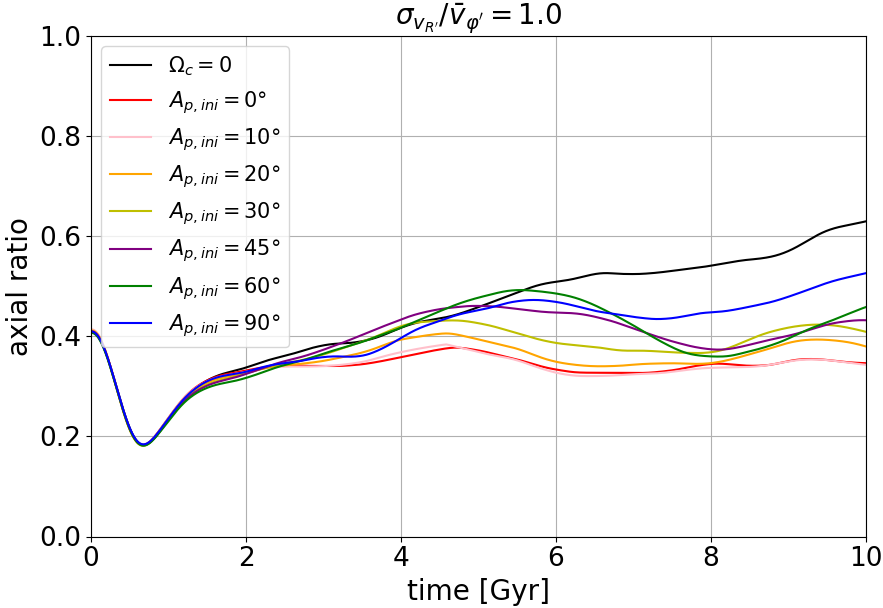}\\
\includegraphics[width=55mm]{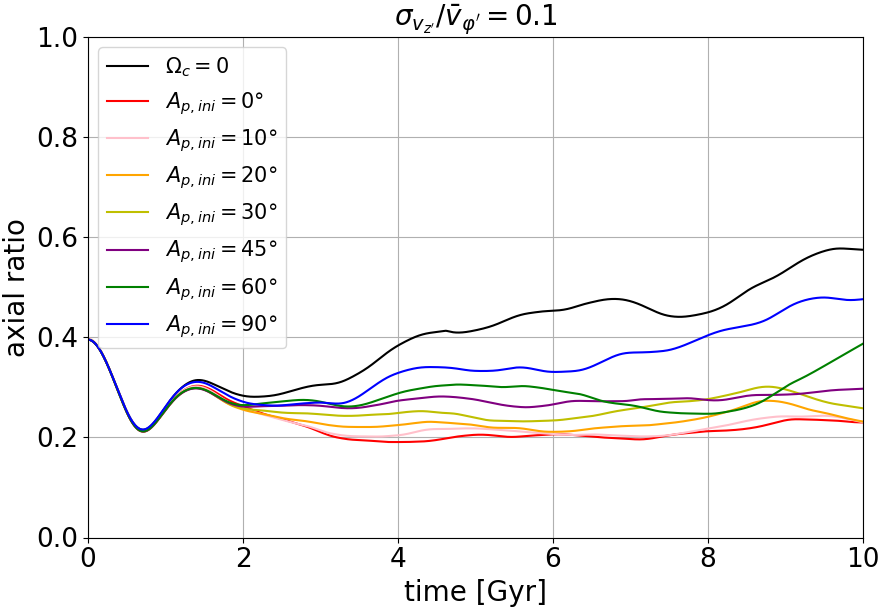}
\includegraphics[width=55mm]{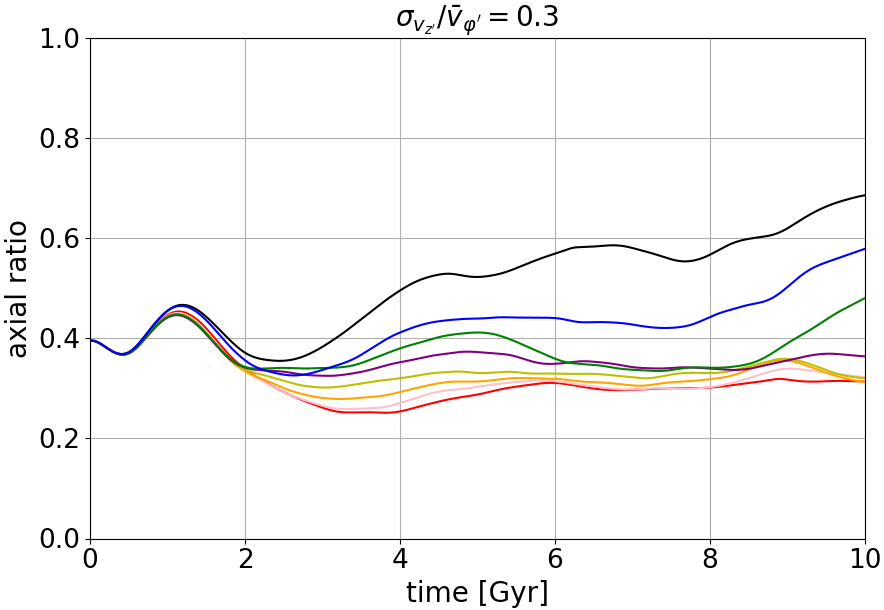}
\includegraphics[width=55mm]{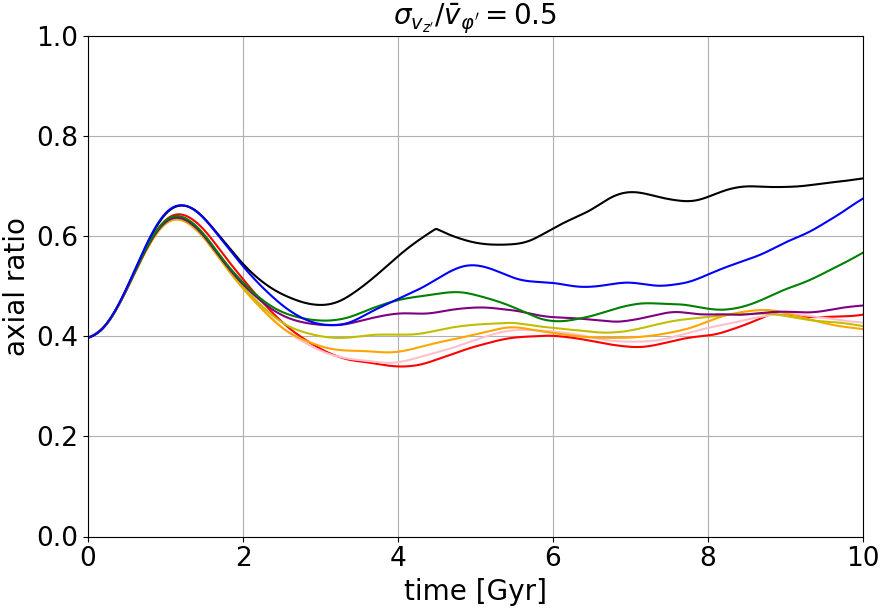}\\
\includegraphics[width=55mm]{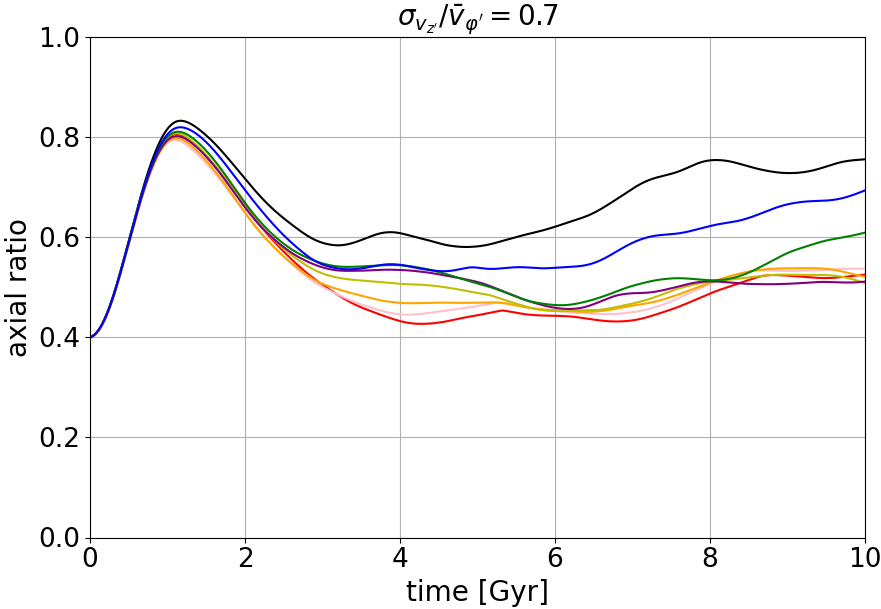}
\includegraphics[width=55mm]{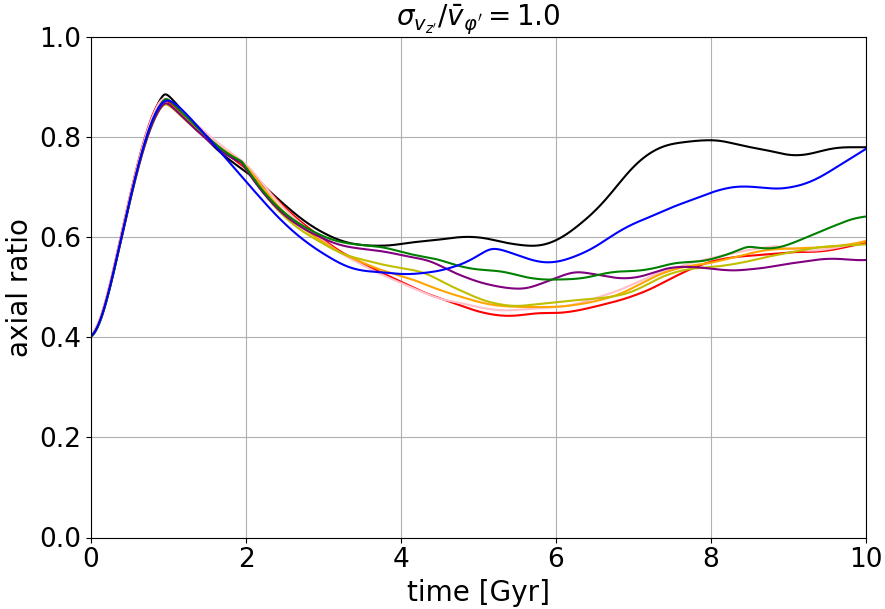}
\caption{The evolution of the axial ratio using collisionless particles.
The most upper panel shows the result starting from the initial state with only the circular velocity, $v_{R'}=v_{z'}=0$.
Panels on the second row show the case including the radial velocity dispersion $\sigma_{v_{R'}}$, 
and panels on the lower two rows show the case including the vertical velocity dispersion $\sigma_{v_{z'}}$.
The colors of the graphs indicate the direction of the figure rotation, expressed by the offset from $(\theta, \phi)=(70^\circ, -13^\circ)$ direction, $A_{p, {\rm ini}}$.
The correspondence of the offset to $(\theta_f, \phi_f)$ is described in \S \ref{subsec:data}.
\label{fig:axisratio3}}
\end{figure*}

Figure \ref{fig:axisratio3} shows the evolution of the axial ratio of the particle distribution in these cases with different initial velocity dispersions of particles.
At first, the most upper panel shows the ideal case where the initial particles are in circular orbits.
We find that the model without a figure rotation shows the largest axial ratio in this ideal initial condition.
This suggests that a figure rotation of a triaxial gravitational potential certainly makes the particle distribution flattened even if the particles are collisionless.
Especially, if the initial offset between the directions of the figure rotation and the particle orbits (expressed as $A_{p, {\rm ini}}$) is smaller than $45^\circ$, the axial ratio is confined to be as small as 0.2 compared to the initial value of 0.4.

Next, we investigate the cases when the initial states of particles include radial velocity components as shown in the second upper row of Figure \ref{fig:axisratio3} with $\sigma_{v_R'}\neq0$.
We find that the axial ratio is slightly larger than the case without radial velocities, but as long as the initial offset $A_{p, {\rm ini}}$ is smaller than $45^\circ$, the axial ratio keeps smaller than $\sim0.4$ even if the radial velocity is large as $\sigma_{v_{R'}}/\Bar{v}_{\phi'}=1$.
Therefore, it is suggested that the radial velocity of particles is not important in the stability of a satellite plane.

Finally, the lower two rows of Figure \ref{fig:axisratio3} show the cases where the initial states include vertical velocity components.
It is clear from these plots that the offset and the vertical velocity are required to be smaller than $A_{p, {\rm ini}}\lesssim45^\circ$ and $\sigma_{v_{z'}}/\Bar{v}_{\phi'}\lesssim0.5$, respectively, to keep an axial ratio as small as $\sim0.4$.
In Appendix A.1, we have shown that the offset between the directions of the figure rotation and the satellite orbits needs to be small, and here we find that this angle offset can be as large as $45^\circ$.

It is shown that the collisionless cases require the more strict conditions to keep the flatness of particle (or satellite) distribution than the fluidal cases.
At first, the case of a triaxial potential leads to a small axial ratio if the offset between the directions of the figure rotation and the orbital plane is small, whereas it is not the case in an almost spherical potential under the same condition, as shown in Figure \ref{fig:axisratio2}.
This difference can be understood from the fact that the effect of pulling particles back to the stable orbital plane in a triaxial potential is stronger than in a spherical potential.
This is because, under a spherical potential, any plane passing through the center of the potential is a stable plane for an individual orbit, so particles in each panel can move independently without any restriction.
On the other hand, in a triaxial potential, a stable orbit occurs only in a certain plane, so particles move toward the plane and their orbits tend to be aligned along it.
However, if the offset between the above-mentioned two directions or the vertical velocity of particles is large, the particle distribution does not reach a flattened structure due to the large randomness of the velocity distribution.
Therefore, all of the shape of the potential, the direction of the figure rotation, and the vertical velocity of particles are important in forming the flattened distribution of particles.

\begin{figure*}[tp]
\centering
\includegraphics[width=65mm]{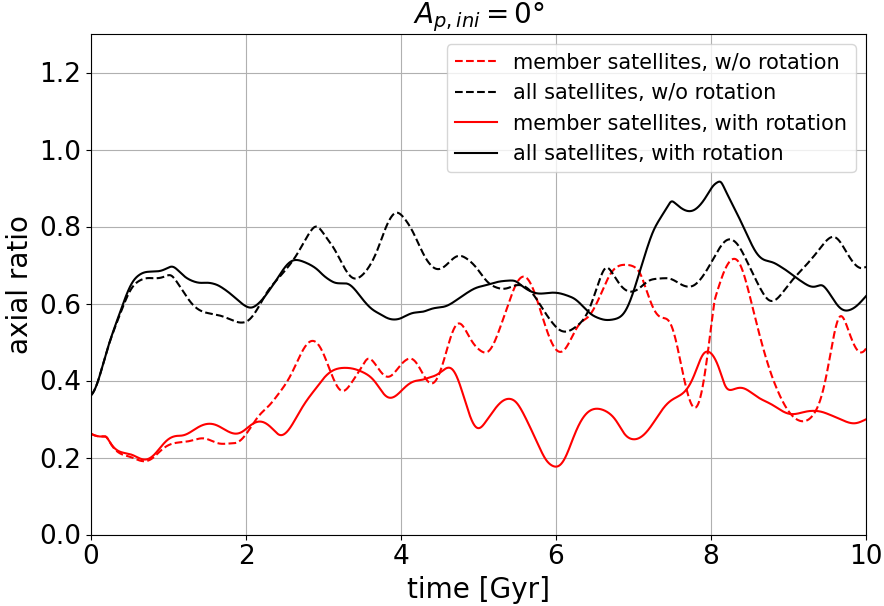}
\includegraphics[width=65mm]{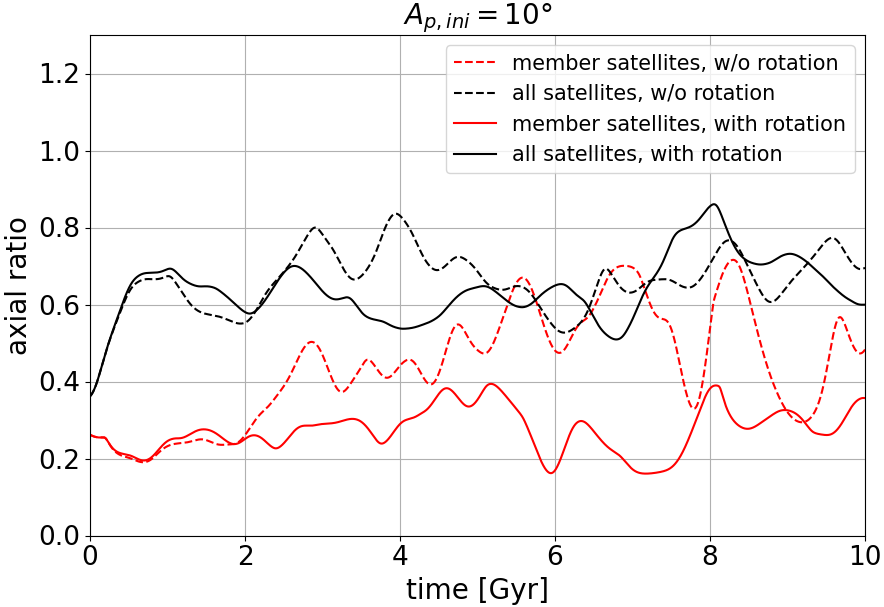}\\
\includegraphics[width=65mm]{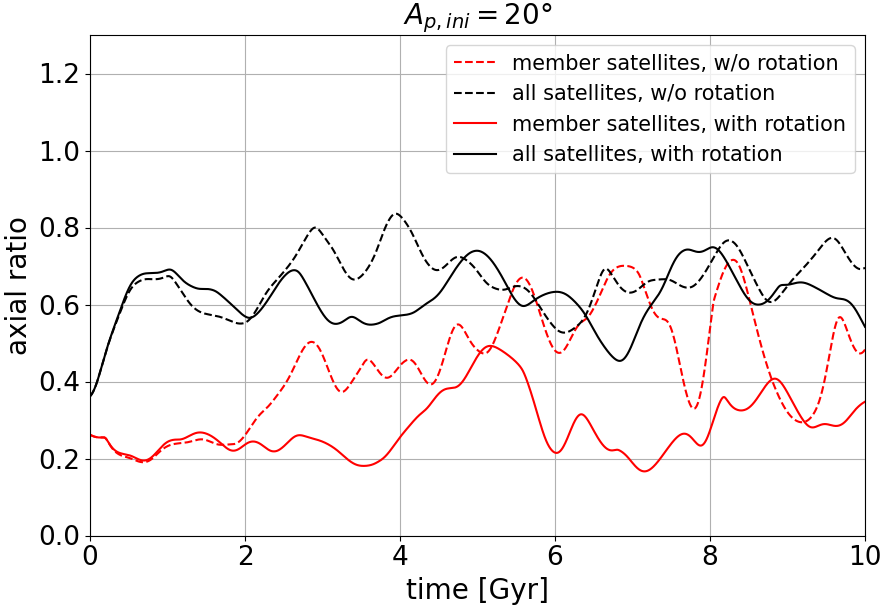}
\includegraphics[width=65mm]{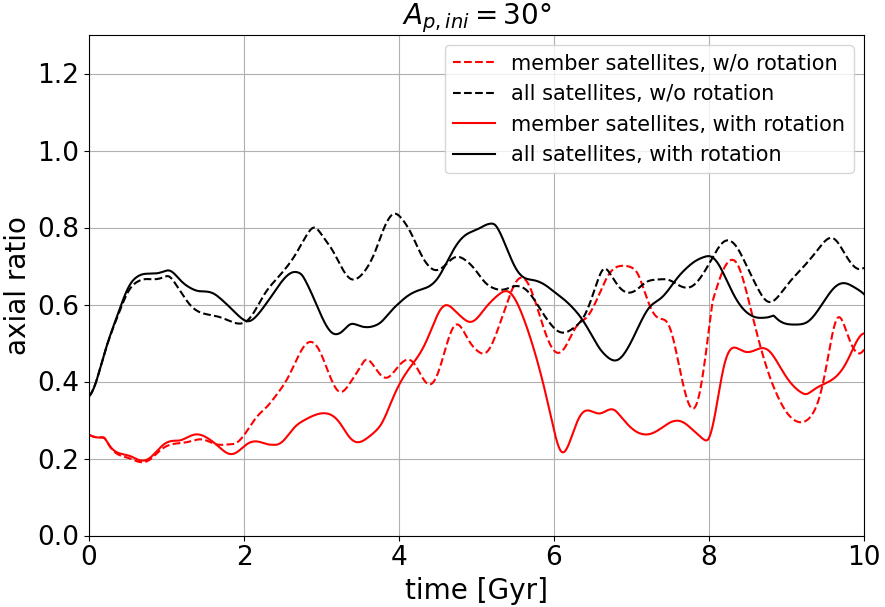}
\caption{The evolution of the axial ratio of the observed satellite distribution.
The black and red graphs show the evolution of the distribution of all satellites and only member satellites, respectively.
The solid and dashed graphs correspond to the evolution under the potential with and without the figure rotation.
Each panel shows the case with a different direction of the figure rotation, labeled by the offset $A_{p, {\rm ini}}$.
\label{fig:axisratio_obs}}
\end{figure*}

These properties of stabilized orbits obtained here are consistent with and the extension of the results of the previous work presented in \citet{valluri2021}.
They found that a figure rotation of a host galaxy shifts an inclination of a tube orbit in a certain amount which is stable if the potential has no figure rotation. 
We note that the inclination of an orbit after this shift does not necessarily coincide with that of the flattened orbital distribution of particles obtained in our work:
since an individual orbit is stabilized by shifting its inclination angle in a specific amount under a figure rotation of gravitational potential, 
the ensemble of these stabilized orbits shows the distribution of inclination angles, which depend on the initial state of particles as well as the property of a figure rotation, such as its rotational direction and speed, as shown in the current work.

\section{Discussion: Origin of the Observed Satellite Plane} \label{sec:dscs}

Our results suggest that the stable orbital plane can occur in a rotating triaxial potential.
To form and keep the stable orbital plane, the direction of the axis of the figure rotation with respect to the normal line of the satellite plane is confined to be within $45^\circ$.
Additionally, the velocity distribution of satellites should satisfy the condition $\sigma_{v_{z'}}/\Bar{v}_{\phi'}\lesssim0.5$.

To assess if the latter condition is satisfied in the observed MW satellites or not, we analyze their velocity distribution and find $\sigma_{v_{z'}}/\Bar{v}_{\phi'}=2.32$ \footnote{We remove the satellites whose apocentric radius is larger than 500~kpc in following calculations. These satellites go extremely far from others and distort the shape of the distribution. The applicable satellites are Columba I, Eridanus II, Leo I, Leo V, Phoenix, and NGC 6822}, i.e., larger than the upper limit, 0.5.
It suggests that the flatness of the observed satellite plane is not maintained by the figure rotation with $\Omega_c$ of $\sim1.0~{\rm km~s^{-1}~kpc^{-1}}$.
Actually, we calculate the orbits of the observed satellites and the evolution of the axial ratio under our triaxial potential, as shown in Figure \ref{fig:axisratio_obs} by black lines.
It follows that regardless of the existence of the figure rotation or its direction, the observed axial ratio $\sim0.4$ is not maintained.
A faster figure rotation may keep the observed flatness, but such a figure rotation is unrealistic \citep{bailin2004, bryan2007}.

On the other hand, \citet{taibi2023} recently presented the member satellites belonging to the planar structure based on their orbital poles. 
Following this work, we identify 16 members \footnote{Boo\"{o}tes III, Carina, Carina III, Crater II, Draco, Fornax, Grus II, Horologium I, Hydrus I, Phoenix II, Reticulum II, Sculptor, Tucana IV, Ursa Minor, LMC, SMC} from our sample.
We find that their velocity distribution has $\sigma_{v_{z'}}/\Bar{v}_{\phi'}=0.52$, so the members almost satisfy the condition to keep their flatness.
Figure \ref{fig:axisratio_obs} also shows the evolution of the axial ratio of the member satellites by red lines.
It follows that the flatness of the member satellites is maintained if the figure rotation exists and the offset between the directions of the axis of the figure rotation and the normal line of the satellite plane is confined to $A_{p, {\rm ini}}<20^\circ$.
This condition about the offset is more strict than the case using our $N=1,000$ test particles ($A_{p, {\rm ini}}<45^\circ$) because of only the small number of the member satellites ($N=16$), so that the flatness of their global distribution is sensitive to each of these satellites.
We thus conclude that the condition of $A_{p, {\rm ini}}<20^\circ$ is necessary to achieve the observed satellite plane.

This alignment of the figure rotation with respect to the planar distribution of satellites is actually implied from recent works on structure formation in the Universe:
it has been suggested that most satellites have accreted along the cosmic filament \citep{libeskind2015, xu2023}.
Furthermore, \citet{lovell2011} showed that this filamentary accretion causes both a figure rotation of a galaxy and a planar structure of satellites along its direction, thereby making the alignment of the two directions inevitable rather than accidental.

On the other hand, some previous works (e.g., Pawlowski et al. 2012a; Shao et al. 2018) already showed that the filamentary accretion of satellites alone cannot explain the flatness of the satellite plane.  
Since the cross-sectional area of the cosmic filament is generally larger than the great circle of the MW’s dark halo \citep{veraciro2011}, the filamentary accretion alone is insufficient to explain the observed flatness of the satellite distribution. 
A more compact accretion is required.

\citet{wang2013} suggested that a group infall is one of the promising hypotheses to explain the satellite plane.
\citet{taibi2023} also suggested that the member satellites have accreted as a group together.
This suggests that the combination of the filamentary accretion and the group infall may explain the existence and stability of the satellite plane.
The filamentary accretion of dark matter induces a figure rotation of the MW's dark halo, whose rotational direction aligns with the direction of the cosmic filament.
On the other hand, a group infall of satellites along the filament causes their anisotropic distribution along it.
As a result, the two directions of the figure rotation and the satellite plane are aligned, so the flatness of the plane is maintained.
Additionally, non-member satellites also accrete along the filament but not as a group, so they distribute somewhat along the satellite plane of members.

The satellite planes of other galaxies in the Local Group also support our suggestion.
\citet{libeskind2019} showed that the directions of most satellite planes align with the direction of the cosmic large-scale structure approximately.
Thus, it is supported that accretions along the cosmic large-scale structure are related to the construction of satellite planes.

Finally, we point out some flaws of our suggestion. 
It is not clear how compact a group infall is needed to explain the observed flatness of the satellite plane or to cause the velocity distribution $\sigma_{v_{z'}}/\Bar{v}_{\phi'}\lesssim0.5$.
Also, whether a group infall satisfying sufficient compactness can actually be realized should be investigated to show the validity of our suggestion. 
It is also unclear about the status of non-member satellites in the observed satellite system, which may be accreted along the cosmic filament as well.
They may have been accreted just recently, but as shown by black lines in Figure \ref{fig:axisratio_obs}, the flatness of all observed satellites does not hold for even 1 Gyr regardless of the existence of the MW’s figure rotation. 
It is inferred that all satellites have accreted within 1 Gyr or satellites have accreted continuously every 1 Gyr. 
While the former hypothesis appears unrealistic, how valid the latter hypothesis is, including the flatness of every accretion event, needs to be clarified. 
To resolve these questions and understand the origin of the satellite plane in the framework of the $\Lambda$CDM model, 
high-resolution, cosmological simulations for the MW-sized galaxies are needed to investigate the formation of the satellite plane considering the filamentary accretion, group infall of satellites, and the figure rotation. 

\section{Conclusion} \label{sec:cncl}
It has been pointed out that satellite galaxies around a host galaxy tend to distribute along a large-scale plane perpendicular to its stellar disk.
Since this planar structure is hardly explained by the $\Lambda$CDM model, this is called the satellite plane problem and is regarded as one of the strongest tensions against the standard cosmological model.

In this work, we have investigated a mechanism for forming and stabilizing a satellite plane based on the figure rotation of the gravitational potential of a host galaxy.
At first, we examined the distribution of a stable orbital plane in a potential with a figure rotation by using fluidal particles.
Next, we have explored the conditions under which satellites form a planar structure and its flatness is maintained by using collisionless particles.
As a result, we have obtained the following results.

(i) In a rotating, nearly spherical potential, a stable orbital plane is precessing around the axis of the figure rotation.
The frequency of the precession is proportional to the pattern speed of the figure rotation.
Its amplitude coincides with the offset between the directions of the axes of the figure rotation and the initial tube orbits of particles.

(ii) In a rotating triaxial potential, a stable orbital plane occurs perpendicular to the axis of the figure rotation if the axis of the figure rotation coincides with the axis of the initial tube orbits of particles.
This coincidence is permitted within $45^\circ$, but this value may depend on the number of satellites.

(iii) For collisionless particles, they can remain around the stable plane and keep their flattened structure if their velocity distribution satisfies the condition of $\sigma_{v_{z'}}/\Bar{v}_{\phi'}\lesssim0.5$, where $v_{z'}$ is the velocity component perpendicular to the planar structure and $v_{\phi'}$ is the azimuthal velocity on the plane.

(iv) While all the MW's satellites found so far do not satisfy the above condition about the velocity distribution, the member satellites of the MW's satellite plane identified in \citet{taibi2023} satisfy it.
For these members, the condition about the offset between the axis of the figure rotation and the axis of the initial tube orbits is more strict as $\lesssim20^\circ$.

(v) The coincidence of the directions of the figure rotation and the satellite plane is realized by the filamentary accretion.
Additionally, combining the concept of the group infall with the filamentary accretion may explain both the formation of the satellite plane and the maintenance of its flatness based on the figure rotation of the host galaxy.

If the satellite plane is explained by this hypothesis, then the figure rotation of the dark halo is tilted to the stellar disk with an inclination of $70^\circ\pm20^\circ$, which is the $\theta$-component of the MW's observed satellite plane (see ${\bm n}$ in Table \ref{tab:obs}).
This different configuration for these components is actually possible because the formation mechanisms of the dark halo and disk are different (e.g., Baptista et al. 2023).
Therefore, it is expected that the observed structure of the MW's satellite plane can constrain the basic property of the MW's figure rotation and its accretion history.

\begin{ack}
We acknowledge support in part from MEXT Grant-in-Aid for Scientific Research (No. JP18H05437 and JP21H05448).
We are grateful to many colleagues for their discussion and suggestions on this work, especially to Yutaka Hiral, Kohei Hayashi, Daniela Carollo, Timothy Beers, Evan Kirby, Borja Anguiano, Teppei Okumura, Keiichi Umetsu, Hei Yin Jowett Chan, Eva K. Grebel, Tadafumi Matsuno, and Marcel S. Pawlowski.

This work has made use of data from the European Space Agency (ESA) mission Gaia, processed by the Gaia Data Processing and Analysis Consortium (DPAC). Funding for the DPAC has been provided by national institutions, in particular the institutions participating in the Gaia Multilateral Agreement.
Full Gaia data credits: https://www.cosmos.esa.int/web/gaia-users/credits

\end{ack}

\appendix

\section{Fluidal Particle Case}
In this appendix, we explore general stable orbits in a rotating system as mentioned in \S \ref{sec:intro}.
We attempt to find closed obits by introducing fluidal interaction among particles, 
i.e., pressure and viscosity effects, so that mutual crossing of orbits is suppressed. 
Based on the results under the fluidal condition, we investigate the stability of the satellite plane as the stable orbital plane in the main text.

\subsection{Calculational Conditions for Fluidal Particles}
To include fluidal effects among particles, we adopt the smoothed particle hydrodynamics (SPH) method to calculate the motion for these fluidal particles. 
In this calculation, we assume the mass of particles is that of hydrogen, $m=m_H=1.67\times10^{-27}~{\rm kg}$.

The equations of motion for fluidal particles are expressed as follows \citep{habe1985, habe1988}. 
\begin{align}
\frac{d{\bm x}_i}{dt} &= {\bm v}_i - {\bm \Omega} \times {\bm x}_i
\label{eq:EOM_x2}\\
\frac{d{\bm v}_i}{dt} &= -{\bm \nabla}\Phi({\bm x}_i) -\frac{1}{\rho_{\rm loc}({\bm x}_i)} {\bm \nabla}[P({\bm x}_i)+\nu({\bm x}_i)] - {\bm \Omega} \times {\bm v_i}
\label{eq:EOM_v2}
\end{align}
Compared with Equation \eqref{eq:EOM_x0} and \eqref{eq:EOM_v0}, the pressure term $P({\bm x})$ and the viscosity term $\nu({\bm x})$ are added.

$\rho_{\rm loc}({\bm x})$ is the local density defined by
\begin{equation}
\rho_{\rm loc}({\bm x})=\sum^N_{j=1}mW({\bm x}-{\bm x}_j).
\label{eq:local_density}
\end{equation}
$W({\bm x})$ is the broadening function and is defined as a Gaussian form,
\begin{equation}
W({\bm x})=\frac{1}{\pi^{3/2}\sigma^3}\exp\left({-\frac{|{\bm x}|^2}{\sigma^2}}\right).
\label{eq:broadening_function}
\end{equation}
We assume $\sigma=25.0~{\rm kpc}$ for our calculation over the range of $r<200~{\rm kpc}$ from the center of the potential. 
This is comparable to the SPH simulation by \citet{habe1985} and \citet{habe1988}, where the value of $\sigma$ is set as $250~{\rm pc}$ over the calculated radial range of several kpc, 
thereby our calculation is a scaled-up version of their model by two orders of magnitude.

$P({\bm x})$ is the pressure term.
We assume the barotropic and isothermal gas, thus, 
\begin{equation}
P({\bm x}) = \frac{k_BT}{m}\rho_{\rm loc}({\bm x})
\label{eq:pressure}
\end{equation}
where $k_B$ is the Boltzmann constant and we assume $T=10^4~{\rm K}$ as the temperature of particles \citep{habe1985}.

$\nu({\bm x})$ is the viscosity term given as.
\begin{align}
\nu({\bm x}) &= \sigma^2\rho_{\rm loc}({\bm x})~[{\bm \nabla}\cdot{\bm v}({\bm x})]^2~ &( {\rm if}~{\bm \nabla}\cdot{\bm v}({\bm x})<0 ) \nonumber \\
&= 0~ &( {\rm otherwise})
\label{eq:viscosity}
\end{align}
The divergence of velocity ${\bm \nabla}\cdot{\bm v}$ is calculated by the relations of ${\bm \nabla}\cdot(\rho_{\rm loc}{\bm v})={\bm \nabla}\rho_{\rm loc}\cdot{\bm v}+\rho_{\rm loc}{\bm \nabla}\cdot{\bm v}$, Equation \eqref{eq:local_density} and
\begin{align}
{\bm \nabla}\rho_{\rm loc}({\bm x}) &= \sum_{j=1}^N m{\bm \nabla} W({\bm x}-{\bm x}_j)
\label{eq:grad_rho}\\
\rho({\bm x}){\bm v}({\bm x}) &= \sum_{j=1}^N m{\bm v}_j W({\bm x}-{\bm x}_j)
\label{eq:rho_vel}\\
{\bm \nabla}\cdot[\rho({\bm x}){\bm v}({\bm x})] &= \sum_{j=1}^N m{\bm v}_j\cdot{\bm \nabla}W({\bm x}-{\bm x}_j)
\label{eq:div_rho_vel}
\end{align}

In this fluidal calculation, we use the first and second initial states shown in the upper and middle rows of Figure \ref{fig:init}.
Because the first initial state is an ideal rotating sphere, it is expected to show a basic orbital structure.
The calculation adopting the second initial state will allow us to understand the behaviors of stable orbits starting from the distribution similar to the observed MW's satellites.

\subsection{Result: Stable Orbital Structure} 
\begin{figure*}[t]
\centering
\includegraphics[width=80mm]{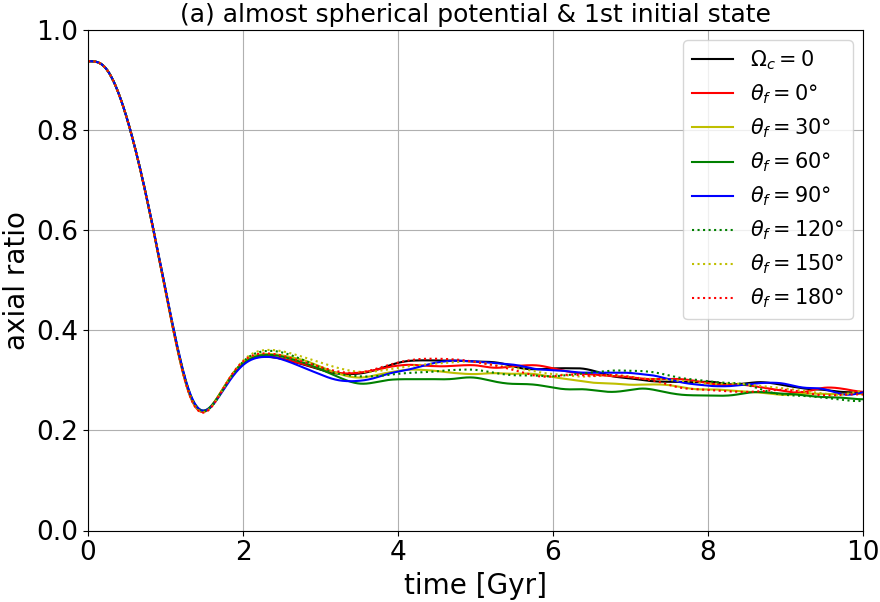}
\includegraphics[width=80mm]{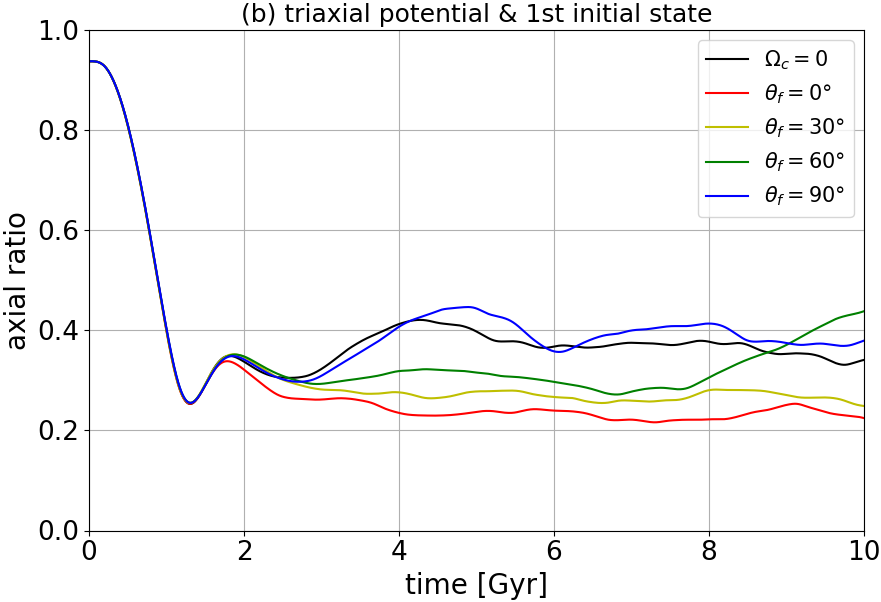}\\
\includegraphics[width=80mm]{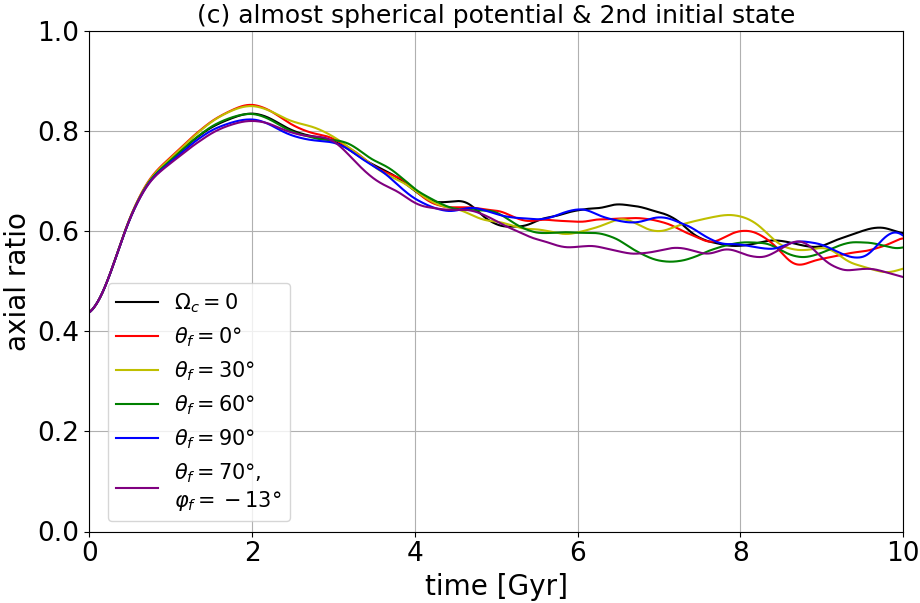}
\includegraphics[width=80mm]{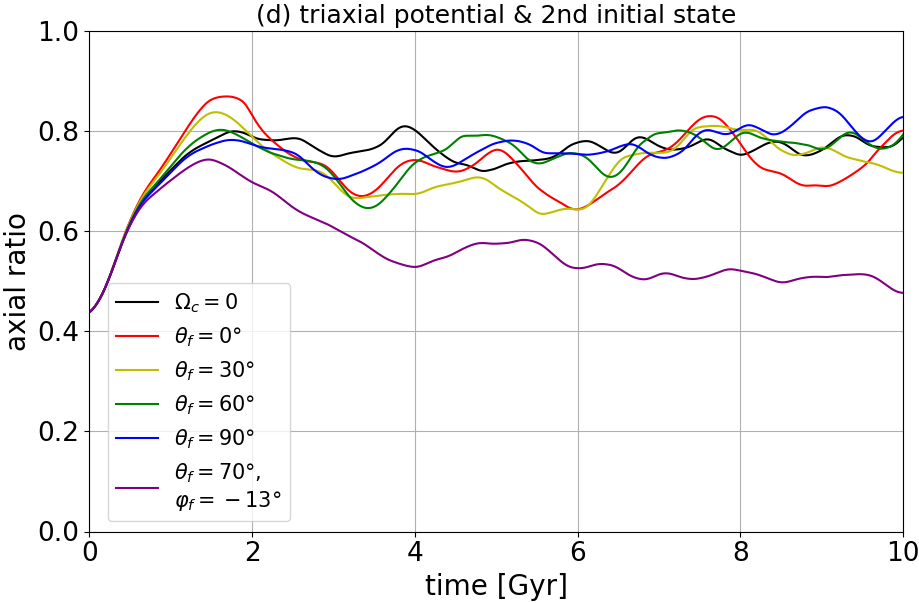}
\caption{The evolution of the axial ratio of the distribution using the fluidal particles.
The upper and lower rows show the result with the first and the second initial conditions, respectively.
The left and right columns show the cases with the oblate-like and the MW-like potential, respectively.
\label{fig:axisratio1}}
\end{figure*}

Figure \ref{fig:axisratio1} shows the evolution of the axial ratio of the particle distribution in four cases, i.e., for the two potential models (the nearly spherical, oblate-like potential: left panels, the triaxial potential: right panels) and the two initial states (the first state: upper panels, the second state: lower panels).
We find that the particle distribution with the first initial state is settled into a flattened shape in both potentials. 
In an almost spherical potential model in panel (a), all curves mostly overlap with each other reaching a small axial ratio, $\sim0.28$ at 10~Gyr.
It is suggested that this flattened distribution is caused by the fluidal terms, not by the figure rotation.
In a more triaxial potential model in panel (b), the axial ratio also decreases to $0.2-0.4$ with slight fluctuation depending on the direction of the figure rotation. 

The reason why the particle distribution gets flattened is that there is no expulsion force against gravity along the axis of a tube orbit, 
whereas the centrifugal force prevents the system from collapsing perpendicular to the axis.
Thus, the differences occur in the apocentric radii or the scale lengths of the particle distribution along the direction of the axis of tube orbits and its perpendicular direction.
The ratio of the apocentric radii along these two directions reaches $0.2-0.4$ at 10~Gyr. 
Furthermore, the ratio of the scale lengths in the particle distribution also reaches $0.2-0.4$, which thus shows the small axial ratio.

The characteristic spatial lengths are also determined by the distribution of the angular momenta of particles.
Indeed, for the first initial state with short-axis tube orbits (panel(a)), their angular momenta are dominated by their z-component.
In an almost spherical potential, the distribution of angular momentum is conserved, so the particle distribution collapses along the z-axis.
When the system has a figure rotation, all components of the angular momenta are not conserved but the component along the axis of the figure rotation is conserved.
This is why the particle distribution gets flattened with precessing around the rotational axis.

On the other hand, when we adopt the second initial condition, the axial ratio of the particle distribution is not settled into a small value in most cases.
In an almost spherical potential (panel (c)) and triaxial potential (panel (d)), the axial ratio reaches $\sim0.6$ and $\sim0.8$, respectively.
This large axial ratio may be due to the large velocities of particles so that it is difficult to form a flattened structure of a stable orbital plane even including the fluidal terms.

\begin{figure*}[t]
\centering
\includegraphics[width=80mm]{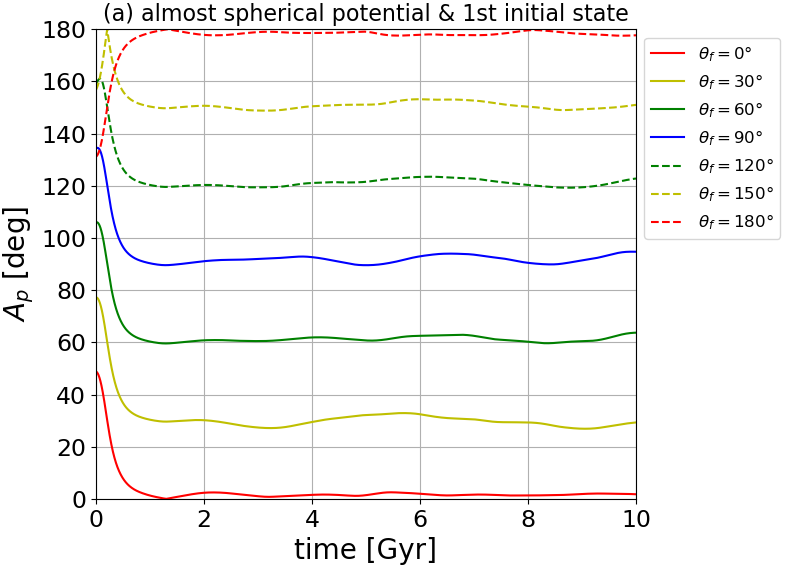}
\includegraphics[width=80mm]{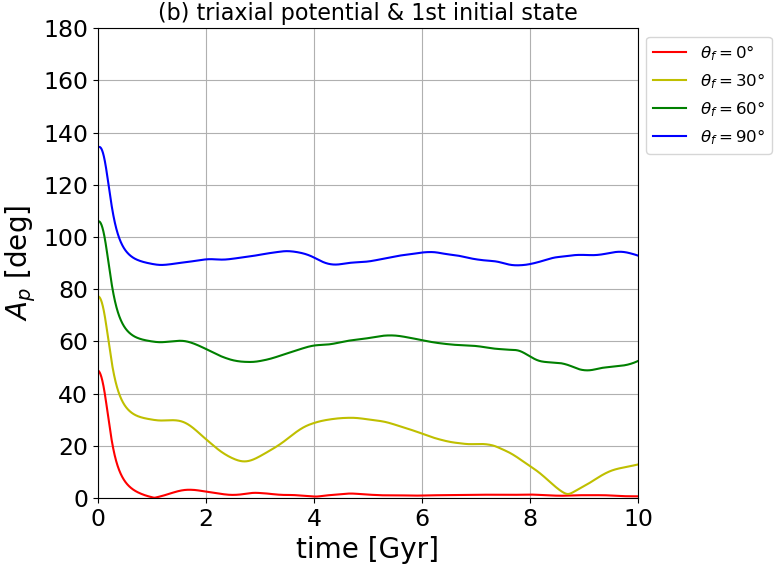}\\
\includegraphics[width=80mm]{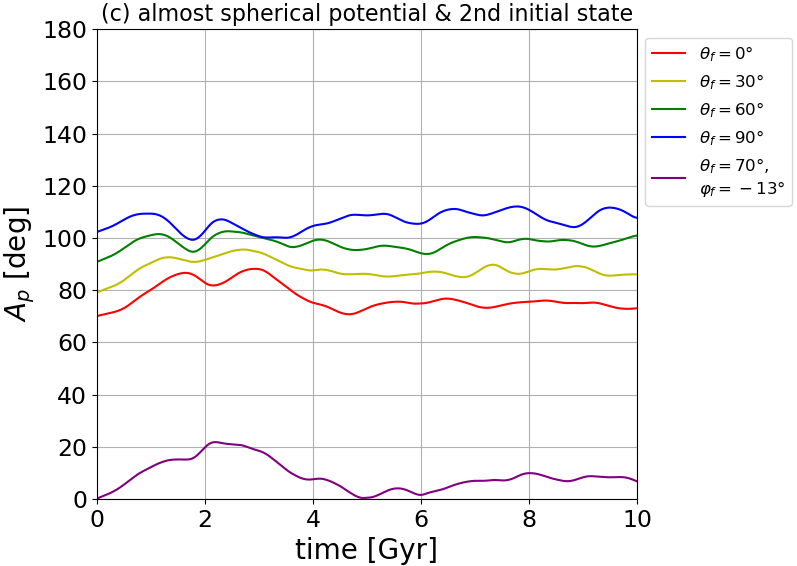}
\includegraphics[width=80mm]{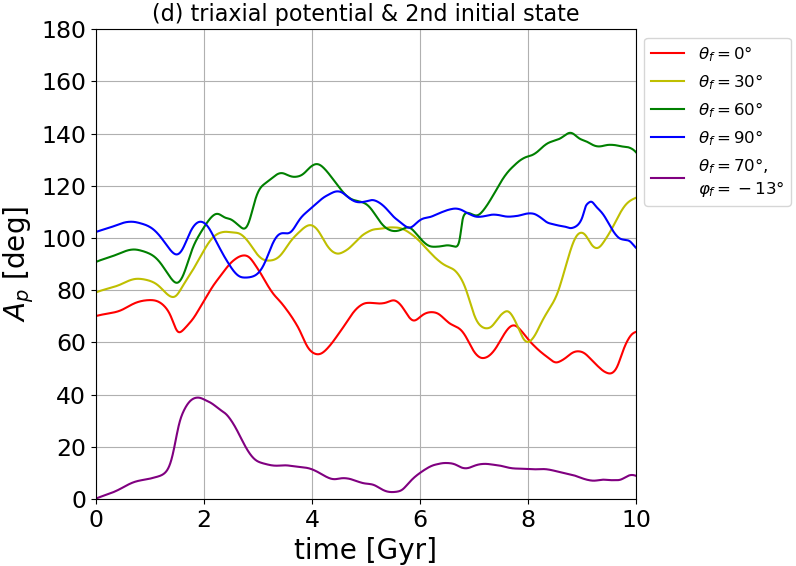}
\caption{The evolution of the direction of the particle distribution, expressed by $A_p\equiv\arccos{({\bm \Omega}/\Omega_c\cdot{\bm u}_3)}$.
As in Figure \ref{fig:axisratio1}, the left and right columns show the cases with the almost spherical (i.e., oblate-like) and the triaxial (i.e., MW-like) potential, respectively.
The upper and lower rows show the result with the first and the second initial conditions, respectively.
\label{fig:amplitude1}}
\end{figure*}

However, there is an exceptional case in a triaxial potential model in panel (d);
when the axis of the figure rotation points to $(\theta_f, \phi_f) = (70^\circ, -13^\circ)$, the axial ratio gets small $\sim0.5$ at 10~Gyr (purple line in panel (d)).
This exception corresponds to the case when the axis of the figure rotation coincides with the normal line of the planar structure.
Except for this special case, an initially planar distribution of fluidal particles is dispersed and gets spherical.

In the cases where we adopt the tilted tube orbit as the initial state or a triaxial potential,
the conservation of the angular momentum is incomplete.
Therefore, the role of the rotation in supporting the system is less significant, so its shape gets less flattened even if the fluidal terms are considered.
Especially, by combining the tilted initial state and the more triaxial potential (panel (d)), the angular momentum is not conserved at all 
and the particle distribution gets spherical.
However, there is an exceptional case: when the offset between the axes of the initial tube orbit and the figure rotation is small, the distribution of the angular momentum, which is dominant along the rotation axis, is conserved to some extent.
Therefore, the distribution of particles is stabilized in a flattened structure.

Next, we show the direction of the particle distribution in terms of $A_p\equiv\arccos{({\bm \Omega}/\Omega_c\cdot{\bm u}_3)}$.
Figure \ref{fig:amplitude1} shows the evolution of $A_p$ in the four cases: each panel is for the same set of the model parameters in that of Figure \ref{fig:axisratio1}.
It shows that the almost spherical potential causes a constant $A_p$:
$A_p$ is equal to $\theta_f$ in the case with the first initial state in panel (a), and $A_p$ nearly maintains the initial values in the case with the second initial state in panel (c).
The constant $A_p$ means a stable precession around the axis of the figure rotation with a certain amplitude.

We note that in panel (a) the initial $A_p$ is not equal to $\theta_f$ because the initial state is defined independently of the figure rotation.
Since the particles are initially generated randomly in a sphere, the initial ${\bm u}_3$ and $A_p$ are also given randomly, and they have little meaning in the first place.
The behavior that $A_p$ quickly converges to $\theta_f$ is critical.
Additionally, this discussion about the initial $A_p$ is also valid in panel (b) explained later, where we use the same initial state.

The triaxial potential also causes a constant $A_p=\theta_f$ in the first initial state in panel (b), but if $\theta_f$ is as small as $30^\circ$, $A_p$ is damped and reaches $\sim0^\circ$.
On the other hand, in the case with the triaxial potential and the second initial state (panel (d)), most $A_p$ changes drastically with time.
However, the case with $(\theta_f, \phi_f) = (70^\circ, -13^\circ)$ is still an exception (purple line): $A_p$ keeps around $0^\circ$ after $\sim3$~Gyr.
Therefore, it is suggested that a rotating triaxial potential forms a stable orbital plane perpendicular to the axis of the figure rotation if the offset between the directions of the figure rotation and the initial tube orbit is sufficiently small.

\begin{figure}[t]
\centering
\includegraphics[width=80mm]{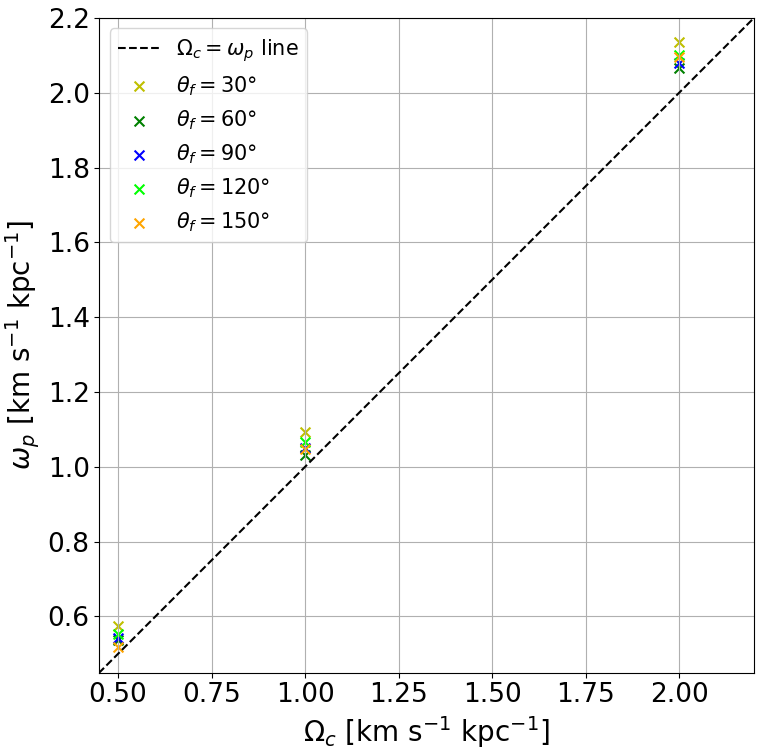}
\caption{The angular frequency of the obtained flattened distribution $\omega_p$ for several $\theta_f$ and $\Omega_c$ cases using the first initial state and the almost spherical potential.
The dashed line is a guideline of $\omega_p=\Omega_c$.
The cases with $\theta_f=0^\circ, 180^\circ$ or $\Omega_c=0$ are omitted because the flattened distributions are almost stable on the equatorial plane and we cannot detect the precession of the particle distribution.
\label{fig:freq10}}
\end{figure}

We analyze the frequency of the precession of the stable orbital plane in the almost spherical potential by changing the pattern speed of the figure rotation.
The frequency is defined by $\omega_p\equiv2\pi/T_p$ and 
$T_p$ is the period of the precession calculated by twice the time it takes for $\theta$ component of ${\bm u}_3$ to go from maximum to minimum or vice versa.
Figure \ref{fig:freq10} shows the relationship between $\Omega_c$ and the mean of $\omega_p$.
It shows that $\omega_p$ is proportional or nearly equal to $\Omega_c$ and mostly independent of $\theta_f$.
Thus, it is concluded that the pattern speed $\Omega_c$ determines the frequency of the precession of the stable orbital plane. 

\section{How to generate the second initial state based on the observed satellite distribution}
In this part, we describe the way to generate our second initial state based on the observed satellites whose number is 49.
We use the satellite distribution in six-dimensional phase space, i.e., defined with the position and velocity, $(x_1, x_2, x_3, x_4, x_5, x_6)\equiv(x, y, z, v_x, v_y, v_z)$.
Additionally, we use an upper index $k$ as the number of a satellite $(k=1-49)$.
For example, $x_5^k$ shows the $y$-component of the velocity of the $k$-th satellite.
Then, the coordinates of the $k$-th satellites in the phase space are expressed as a vector, as ${\bm x}^k=(x_1^k, x_2^k, x_3^k, x_4^k, x_5^k, x_6^k)$.

Since the observed satellite distribution in the spatial space is tilted for the coordinate axes as shown in Figure \ref{fig:observed_satellite}, 
the fitted ellipsoid in the phase space is not along the coordinate axes.
Thus, before we generate test particles, we convert the coordinates based on the directions of the satellite distribution.
We define the inertial tensor in the phase space as $I_{ij}\equiv\sum_{k=1}^{49}x_i^kx_j^k$ ($i,j=1-6$)
and calculate six eigenvectors of this tensor as ${\bm w}_i$.
Based on these eigenvectors, we obtain the coordinates of satellites as $q_i^k\equiv{\bm x}^k\cdot{\bm w}_i=\sum_jx_j^kw_{i,j}$ ($w_{i,j}$ is the $i$-th component of ${\bm w}_{j}$).

We generate test particles along this coordinate.
We calculate the mean and the standard deviation of $q_i$ of 49 satellites, expressed as $\mu_{q_i}$ and $\sigma_{q_i}$, respectively.
Based on these values, we make the distribution function of $q_i$ as a Gaussian form, 
\begin{equation}
G(s; \mu_{q_i}, \sigma_{q_i})=\frac{1}{\sqrt{2\pi\sigma_{q_i}}}\exp{\frac{(s-\mu_{q_i})^2}{2\sigma_{q_i}^2}}.
\end{equation}
We generate 1,000 test particles, $q_i^l$ ($l=1-1,000$) following this distribution function.

We convert back to the original coordinate as $x_i^l\equiv\sum_jq_j^lw_{i,j}^{-1}=\sum_jq_j^lw_{j,i}$.
We note that when we regard ${w_{i,j}}$ as a tensor, it is an orthogonal matrix so $w_{i,j}^{-1}=w_{j,i}$.

When we generate these test particles, we constrain that their apocenters are smaller than $r=500~{\rm kpc}$
We calculate the orbit of each $l$-th test particle in our oblate-like potential defined in \S \ref{subsec:eq}.
If the apocenter is larger than $r=500~{\rm kpc}$, we remove the particle and generate an alternative particle.
This constraint is required in order to reduce the effect of such an outlier particle located extremely far from others, which irregularly distorts the shape of a distribution.
The threshold value of $500~{\rm kpc}$ is determined based on the observed satellites.
The apocenters of most observed satellites in the oblate-like potential are smaller than $400~{\rm kpc}$, but few satellites have larger apocenter than $500~{\rm kpc}$.
These satellites are almost not bound for the MW's gravitational potential and distort the evolution of the axial ratio of the satellite distribution.
Thus, we define the threshold value of $500~{\rm kpc}$.

After we generate 1,000 test particles $x_i^l$, we tilt the particle distribution so that its short-axis direction coincides with the normal line of the observed satellite plane, ${\bm n}=(\theta, \phi)=(70^\circ, -13^\circ)$.
The short-axis direction is obtained by the spatial inertial tensor ${S_{ij}}$ and its third eigenvector ${\bm u}_3$ as explained in \S \ref{sec:obs_vpos} but using test particles.

Finally, we align the rotational velocity components of test particles because it is irrational that fluidal particles rotating in opposite directions are mixed.
If a particle rotates clockwise around ${\bm n}$, we change the sign of the rotational velocity component so that it rotates anti-clockwise.

As a result, the distribution of these test particles has a flattened spatial distribution with the axial ratio $\lambda_3/\lambda_1=0.438$.
Although this value is different from that of the observed satellites, $\lambda_3/\lambda_1=0.367$, this difference may be due to the difference between the number of observed satellites ($N=49$) and test particles ($N=1,000$).
Because the number of particles increases from the observed satellites, the generated distribution expands and gets a slightly larger axial ratio than the observed distribution.
We judge that this flatness difference does not have much of an impact on the result.

\end{document}